\documentclass[%
twocolumn,
nofootinbib,
amsmath,amssymb,
balancelastpage,
aps,
prd,
floatfix,
]{revtex4-2}

\usepackage{graphicx}
\usepackage{dcolumn}
\usepackage{bm}
\usepackage{siunitx}
\usepackage{physics}
\AtBeginDocument{\RenewCommandCopy\qty\SI}

\usepackage{easyReview}
\usepackage{slashed}
\usepackage{mathtools}
\usepackage{xspace}
\usepackage{comment}
\usepackage{braket}
\usepackage{float}
\usepackage{fontawesome}
\usepackage{booktabs}
\usepackage{anyfontsize}
\usepackage{listings}
\usepackage{inconsolata}
\usepackage{multirow}
\usepackage{tikz}
\usepackage{xpatch}
\xpatchbibdriver{misc}
  {\printtext[parens]{\usebibmacro{date}}}
  {\iffieldundef{year}
    {}
    {\printtext[parens]{\usebibmacro{date}}}}
  {}
  {\typeout{Removed date because none provided}}

\tikzset{every picture/.style={line width=0.75pt}}
\usetikzlibrary{calc}
\usetikzlibrary{decorations.pathmorphing}

\definecolor{linkcolor}{rgb}{0.0, 0.28, 0.67}
\definecolor{userInputColor}{HTML}{4287f5}
\definecolor{userCallColor}{HTML}{fcba03}
\definecolor{codeCallColor}{HTML}{039c33}
\usepackage[dvipsnames]{xcolor}

\usepackage[
    colorlinks=true,
    urlcolor=linkcolor,
    anchorcolor=linkcolor,
    citecolor=linkcolor,
    filecolor=linkcolor,
    linkcolor=linkcolor,
    menucolor=linkcolor,
    linktocpage=true,
    pdfproducer=medialab,
    pdfa=true
]{hyperref}

\DeclareSIUnit\erg{erg}
\DeclareSIUnit\year{yr}
\DeclareSIUnit\jansky{Jy}
\DeclareSIUnit\dex{dex}
\DeclareSIUnit\deg{deg}
\DeclareSIUnit\angstrom{\text{Å}}
\DeclareSIUnit\eV{e\kern-.05em V}
\DeclareSIUnit \parsec {pc}
\DeclareSIUnit[]\msol
{\text{\ensuremath{M_{\odot}}}}
\DeclareSIUnit\beam{beam}
\graphicspath{ {./images/} }

\defcitealias{Baker:2025bnz}{Paper~II}

\newcommand{\githubicon}{\href{https://github.com/bakerem/dark_photons_radio_sky/tree/v1.0.0}{\faGithub}\xspace}
\newcommand{\nblink}[1]{\href{https://github.com/bakerem/dark_photons_radio_sky/blob/v1.0.0/notebooks_for_paper/#1.ipynb}{\faGithub}}

\begin{document}

\title{Dark Photons in the Radio Sky: I. Resonant Conversions in Halos}
\author{Ethan Baker}
\email{ebaker@bu.edu}
\thanks{ORCID: \href{https://orcid.org/0000-0002-0520-4235}{0000-0002-0520-4235}}
\affiliation{Physics Department, Boston University, Boston, Massachusetts 02215, USA}

\author{Hongwan Liu}
\email{hongwan@bu.edu}
\thanks{ORCID: \href{https://orcid.org/0000-0003-2486-0681}{0000-0003-2486-0681}}
\affiliation{Physics Department, Boston University, Boston, Massachusetts 02215, USA}

\date{\today}
\begin{abstract} 

    Mixing between dark photons and visible photons leads to substantial anisotropies in the cosmic microwave background due to resonant conversions of visible photons into dark photons in baryonic matter found in dark matter halos.  
    In this work, we forecast the sensitivity of the Square Kilometre Array (SKA) to this signal. 
    We find that SKA could be the first experiment to discover dark photons with a mass between $10^{-13}$ and $5\times 10^{-12}$ eV and kinetic mixing parameter $\epsilon$ as small as $\sim 10^{-8}$ by cross-correlating their data with a low-redshift galaxy survey, potentially improving on the sensitivity from a similar analysis using \textit{Planck} data by a factor of 4 in $\epsilon$. 
    This improvement is largely due to an enhancement of the signal at low frequencies and the unique experimental advantages of radio telescopes such as small beam sizes.~\githubicon
\end{abstract}

\maketitle

\section{Introduction}
The dark photon $A'$ is one of the simplest extensions to the Standard Model~\cite{Holdom:1985ag}. 
It is the gauge boson of a new dark $U(1)'$ symmetry, and couples to the Standard Model photon through a kinetic mixing term scaled by a mixing parameter $\epsilon$, enabling $\gamma \leftrightarrow A'$ conversions.
In a plasma, the probability of conversion is enhanced resonantly when the mass of the dark photon $m_{A'}$ is equal to the effective plasma mass $m_\gamma$, which is predominantly determined by the electron number density $n_e$ in the gas. 
This effect allows for the search for $\gamma \leftrightarrow A'$ conversions in a variety of cosmological environments, with $m_\gamma$ ranging from $\qtyrange[range-phrase={~\text{to}~}]{e-15}{e-5}{\eV}$. 
Here, we focus on $\gamma \to A'$ conversions, without assuming that dark photons are the dark matter. 

Previous cosmological searches for $\gamma \to A'$ conversions have focused on the cosmic microwave background (CMB).
Specifically, Refs.~\cite{Mirizzi:2009iz,Kunze:2015noa,Caputo:2020bdy,Caputo:2020rnx,Garcia:2020qrp, Chluba:2024wui,Arsenadze:2024ywr} set limits on the dark photon kinetic mixing parameter $\epsilon$ by searching for spectral distortions of the CMB monopole blackbody spectrum from such conversions. 
References~\cite{Caputo:2020bdy,Caputo:2020rnx,Garcia:2020qrp} found that properly modeling $\gamma \to A'$ conversions in the inhomogeneous plasma of our Universe was crucial to computing this spectral distortion signal accurately.

However, beyond just looking for the sky-averaged disappearance of CMB photons, we can also consider the angular power spectrum sourced by the inhomogeneous nature of $\gamma \to A'$ conversions~\cite{Pirvu:2023lch,McCarthy:2024ozh,Aramburo-Garcia:2024cbz}.
This power spectrum of disappearing photons is determined by the cosmic distribution of free electrons, since $\gamma \to A'$ conversions occur when $m_{A'}^2 = m_{\gamma}^2 \propto n_e$.
Using the halo model and a model for the distribution of baryons within a halo as a function of mass and redshift, the authors of Ref.~\cite{Pirvu:2023lch} showed how to compute the photon disappearance angular power spectrum due to $\gamma \to A'$ in the gas within halos, as well as the cross-correlation of this signal with a galaxy catalog. 
The authors of Ref.~\cite{McCarthy:2024ozh} then performed an actual search for this signal using the \textit{Planck} CMB temperature maps and the \textit{unWISE} galaxy catalog, setting the strongest limits to date on $\epsilon$ for $m_{A'} \sim \qty{e-12}{\eV}$ using the cross-correlation; this signal scales more favorably as $\epsilon^2$, compared to the autocorrelation signal, which scales as $\epsilon^4$.

In this work, we study for the first time $\gamma \to A'$ conversions at radio frequencies. 
We forecast the sensitivity of the Square Kilometre Array (SKA) to dark photons, and find that it will be one of the best ways to search for $A'$ for two main reasons. 
First, the probability of $\gamma \to A'$ conversions occurring is proportional to $\omega^{-1}$, where $\omega$ is the present-day photon energy, and is therefore enhanced by a factor of at least $\sim 30$ at radio frequencies compared to typical CMB frequencies.
Second, radio telescopes such as SKA operate with much smaller beams compared to \textit{Planck}, giving radio telescopes much greater sensitivity to small angular scales, which is highly beneficial to searches for $\gamma \to A'$ conversions. 
\begin{figure*}[t]
    \centering
    \includegraphics[width=\textwidth]{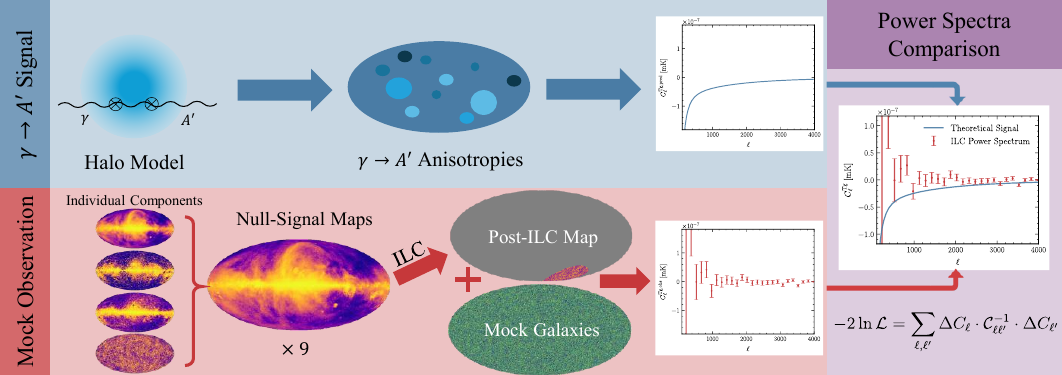}
    \caption{Cartoon representation of the analysis pipeline. We begin with the theoretical prediction of the signal in the halo model. We then cross-correlate this predicted signal with a mock galaxy catalog to get a theoretical power spectrum.
    Next, we simulate astrophysical foregrounds to get nine mock foreground maps at SKA frequencies, which we process with the ILC algorithm. 
    Finally, this null-signal result is compared with the theory prediction using a Gaussian likelihood.}
    \label{fig:pipeline}
\end{figure*}

Our analysis is summarized in Fig.~\ref{fig:pipeline}, and proceeds as follows: 
\begin{enumerate}
\item \textit{Modeling the dark photon signal.}
We model $\gamma \to A'$ conversions in multiple cosmological environments using simulation and analytic methods, as detailed in our companion paper, Ref.~\cite{Baker:2025bnz} (hereafter~\citetalias{Baker:2025bnz}), in order to be sensitive to the largest range of $m_{A'}$ possible.
Here, for clarity, we focus on one representative case: $\gamma \to A'$ conversions in dark matter halos in the range $0.005 \leq z \leq 4$, which we model using the halo-model formalism of Refs.~\cite{Pirvu:2023lch,McCarthy:2024ozh}. 
We compute the angular autopower spectrum of the temperature anisotropies sourced by $\gamma \to A'$ conversions, and its cross-power spectrum with the number density of low-redshift galaxies. 

\item \textit{Producing astrophysical foregrounds-only mock maps.}
In order to forecast the sensitivity of SKA to a potential dark photon signal, we produce nine accurate, mock radio maps at \qtyrange{0.41}{12.53}{\giga\hertz} containing only astrophysical foregrounds and experimental noise, corresponding to the null-signal hypothesis.
The important foregrounds are galactic synchrotron radiation, free-free radiation, spinning dust, and extragalactic radio point sources~\cite{Thorne:2016ifb, Zonca:2021row,Pan-ExperimentGalacticScienceGroup:2025vcd,Gervasi:2008rr,Mittal:2024mzv}.
These mock foreground-only maps form the basis for our forecast analysis in the next step.

\item \textit{Extracting a potential signal and likelihood analysis.}
Given any set of radio maps at different frequencies, we can optimally extract the putative dark photon signal---with a known $\omega^{-1}$ frequency dependence---using a technique known as the internal linear combination (ILC)~\cite{WMAP:2003cmr,Tegmark:2003ve,Delabrouille:2008qd,McCarthy:2023hpa}.
This algorithm constructs the minimum-variance linear combination of the input maps, subject to the constraint that the magnitude of the signal of interest is not reduced, given its known frequency dependence.
In our analysis, we perform the ILC algorithm on the null-hypothesis maps, and compute the autopower spectrum of the resulting post-ILC map as well as its cross-power spectrum with a mock galaxy survey. 
These power spectra are then compared with the theoretical signal to set an upper limit on the sensitivity of SKA to $\gamma \to A'$ conversions using a Gaussian likelihood constructed from these power spectra and the covariance of the post-ILC power spectrum.
\end{enumerate}

In \citetalias{Baker:2025bnz}, we provide a detailed explanation of the dark photon signal, simulation techniques, foreground modeling, the ILC procedure, and statistical methods.
In total, we consider three cosmological environments in which $\gamma \to A'$ conversions can occur: (1) conversions in dark matter halos, the focus of this work, (2) conversions in the intergalactic medium (IGM) during the Epoch of Reionization (EoR) from $5\lesssim z \lesssim 35$, and (3) conversions in the late universe IGM from $0.005 \lesssim z \lesssim 4$. 
Additionally, we determine the sensitivity of 21-cm global signal experiments to $\gamma \to A'$ conversions. 
The remainder of this paper is structured as follows: First, we describe the general formalism of $\gamma \to A'$ conversions and the observational signals they generate. 
Next, we describe our foreground modeling methods. 
We then forecast the sensitivity of upcoming radio experiments to $\gamma \to A'$ conversions and detail our important conclusions regarding the advantages of radio experiments. 
Throughout this work, we use natural units where $\hbar = c  = k_B = 1$. 

\section{Radio Signals of $\gamma \to A'$ Conversions}
Photons and dark photons are described by the Lagrangian
\begin{equation}
    \mathcal{L} \supset -\frac{1}{4}F_{\mu\nu}^2 - \frac{1}{4}F_{\mu\nu}^{\prime2} - \frac{\epsilon}{2}F^{\mu \nu}F'_{\mu\nu} + \frac{1}{2} m_{A'}^2 {A}_\mu^{\prime2}\, ,
\end{equation}
where $m_{A'}$ is the dark photon mass and $\epsilon$ is the kinetic mixing parameter that controls the strength of the interaction~\cite{Holdom:1985ag}. 
The kinetic mixing term causes $\gamma \leftrightarrow A'$ oscillations.
In a plasma, the probability of these conversions is enhanced resonantly when $m_{A'}^2$ equals the effective plasma mass $m_{\gamma}^2(\vec{x})\approx 4\pi \alpha_{\rm EM} n_e(\vec{x})/m_{e}$, where $\alpha_{\rm EM}$ is the electromagnetic fine structure constant, $n_e$ is the free electron number density, and $m_e$ is the electron mass. 

The total probability of conversion along a line-of-sight $\hat{n}$ is given by a sum over all conversions as a photon travels from the last scattering surface to us, 
\begin{equation}
    P_{\gamma \to A'}(\hat{n}) = \sum_i \frac{\pi \epsilon^2 m_{A'}^2}{H(z_i)\omega_0 (1+z_i)^2} \left|\frac{d \ln m_{\gamma}^2(\hat{n},z)}{dz} \right|^{-1}_{z=z_{i}} \, ,
\end{equation}
where $H(z)$ is the Hubble parameter, $z_{i}$ are the redshifts when resonances occur and $\omega_0$ is the present-day energy of the photon~\cite{Mirizzi:2009iz,Caputo:2020rnx}.

These conversions reduce the specific intensity in frequency of the incoming isotropic backlight photons from the CMB along a line of sight by $\Delta I(\omega_0, \hat{n})=-P_{\gamma \to A'}B(\omega_0)$, where
\begin{equation}
    B(\omega_0) = \frac{\omega_0^3}{2\pi^2}\left[\exp\left(\frac{\omega_0}{T_{\gamma,0}}\right)-1 \right]^{-1}
\end{equation}
and $T_{\gamma, 0}=\qty{2.73}{\kelvin}$ is the present-day temperature of the CMB~\cite{Fixsen:2009ug}.
This corresponds to a thermodynamic temperature deficit $\Delta T(\hat{n}) = -P_{\gamma \to A'}(\omega_0, \hat{n})T_{\gamma,0}(1-e^{-x})/x$ with $x\equiv\omega_0/T_{\gamma,0}$.
At the radio frequencies of interest here, this simplifies to $\Delta T(\hat{n}) \approx -T_{\gamma,0 }P_{\gamma \to A'}(\omega_0, \hat{n})$; at frequencies above $T_{\gamma,0}$, however, the temperature deficit is exponentially suppressed. 

To search for this disappearance signal, we begin by constructing two real-space summary statistics: the two-point autocorrelation function $\xi^{\rm TT}$ of $\Delta T$, and the two-point cross-correlation function $\xi^{\rm Tg}$ of $\Delta T$ and the sky-projected fractional overdensity of galaxies, $\delta_g$, defined as:
\begin{align}
    \xi^{\rm TT}(\mu) &\equiv \left\langle \Delta T(\hat{n})\Delta T(\hat{n}')\right\rangle - \left\langle \Delta T \right\rangle^2\, ,\nonumber \\
    \xi^{\rm Tg}(\mu) &\equiv \left\langle \Delta T(\hat{n})\delta_g(\hat{n}')\right\rangle - \left\langle \Delta T(\hat{n})\right\rangle \left\langle \delta_g(\hat{n}) \right \rangle\, ,
    \label{eq:2pcf}
\end{align}
where $\langle \dots \rangle$ indicates the sky average, and $\mu \equiv \hat{n} \cdot \hat{n}'$. We can then define the corresponding angular power spectra,
\begin{equation}
     C^{\rm TX}_\ell = 2\pi\int_{-1}^{1} d\mu \, \xi^{\rm T X}(\mu) P_\ell(\mu) \, ,
\end{equation}
where $\rm X \in \{T, g\}$ and $P_\ell$ is a Legendre polynomial. 
Note that we have assumed isotropy throughout these expressions.  
These are the main observables of interest in this work; we will perform our forecast of SKA's sensitivity to dark photon conversions by comparing the theoretical predictions of these power spectra, $C_\ell^{\mathrm{TX,pred}}$, to $C^{\mathrm{TX,obs}}_\ell$ obtained from mock data. 

\begin{figure*}
    \centering
    \includegraphics[width=\linewidth]{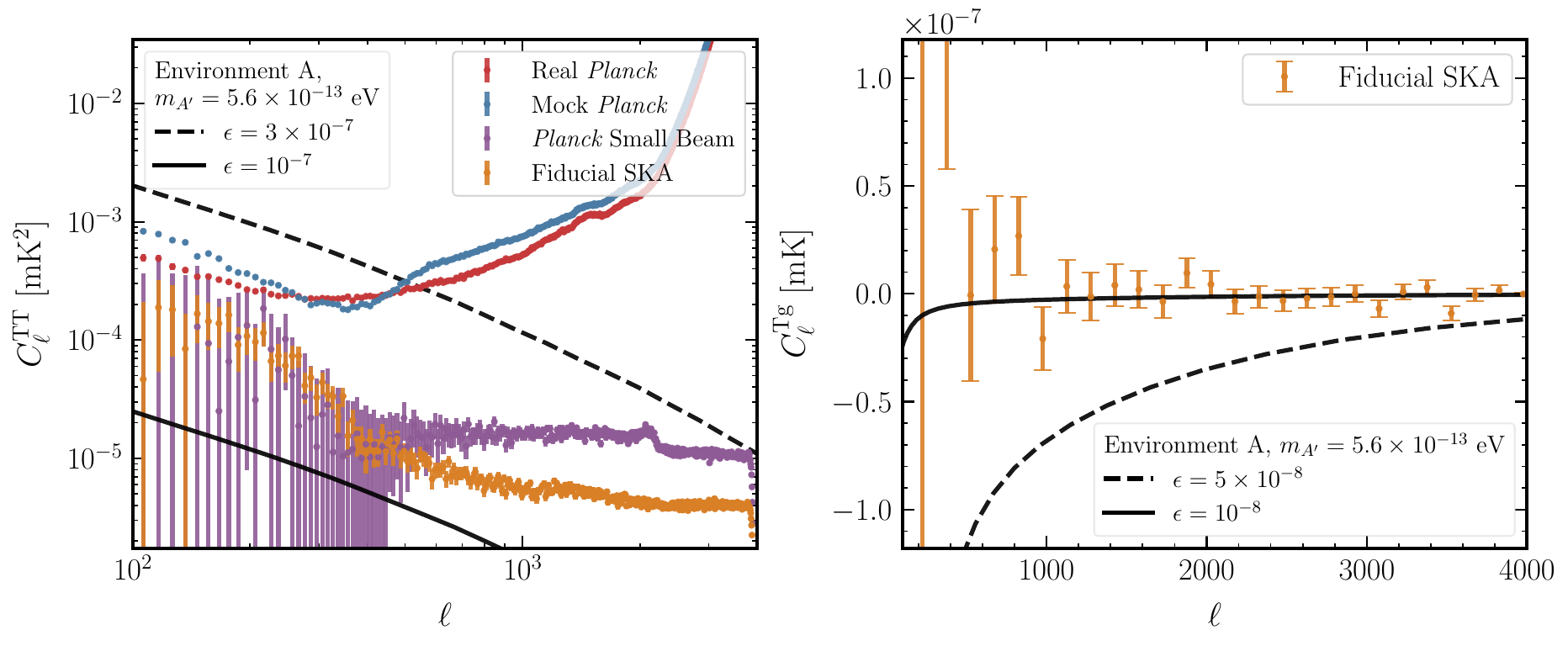}
    \caption{Power spectra of several post-ILC maps compared to the theoretical signal. \textit{Left:} Red and blue lines are the autopower spectra from performing the ILC procedure on real and mock \textit{Planck} maps at the actual beams, respectively. The orange and purple lines are the autopower spectra of our post-ILC fiducial mock radio and mock \textit{Planck} maps, respectively, with a \qty{0.5}{\arcminute} beam. 
    \textit{Right:} Cross-power spectra of our fiducial post-ILC map with our mock $\delta_g$ compared to the theoretical dark photon signal. 
    In both plots, the theoretical signal (in black) is for a dark photon with $m_{A'}=\qty{5.6e-13}{\eV}$.~\nblink{Cl_Comparison}}
    \label{fig:mccarthy_comparison}
\end{figure*}

$C^{\mathrm{TX,pred}}_\ell$ are computed by adopting the methods from Ref.~\cite{McCarthy:2024ozh}, described in more detail in \citetalias{Baker:2025bnz}. 
In summary, we first adopt a halo model, which specifies the number density of halos as a function of their mass and redshift.
Each halo within the distribution is then assigned a baryon distribution from Ref.~\cite{Battaglia:2016xbi}. 
With this, we know how $\gamma \to A'$ conversions should proceed within the gas of each halo. 
Furthermore, the halo model, together with the halo occupation distribution model from Ref.~\cite{Kusiak:2022xkt}, also predicts the distribution of galaxies in the sky.
Combining all of this information, we can compute first the two-point correlation functions in Eq.~\eqref{eq:2pcf}, and then subsequently $C^{\mathrm{TX,pred}}_\ell$.
Representative cases of $C^{\mathrm{TT,pred}}_\ell$ and $C^{\mathrm{Tg,pred}}_\ell$ computed in the halo model are shown as black lines in Fig.~\ref{fig:mccarthy_comparison} for $m_{A'}=\qty{5.6e-13}{\eV}$ at $\omega_0/(2\pi) = \qty{410}{\mega\hertz}$.

\section{Mock Foreground \& Galaxy Modeling}
In order to perform a forecast for SKA, we need to produce foreground-only maps, which correspond to the null-hypothesis that there is no dark photon signal.
The forecast upper limits on $\epsilon$ will then be set by processing these maps with the same pipeline that would be used for real SKA data, computing $C^{\mathrm{TX,obs}}_\ell$ for these maps, and comparing to the theoretical predictions, $C^{\mathrm{TX,pred}}_\ell$.
In this section, we describe our methods for modeling astrophysical foregrounds and producing a mock galaxy catalog that we will then use to forecast the sensitivity of SKA. 
The details of our modeling are described in \citetalias{Baker:2025bnz}.

First, we generate nine mock foreground-only maps in the HEALPix format \cite{healpix} with $N_{\rm side}=2048$ at the SKA observing band centers: \{0.41, 0.56, 0.77, 1.05, 1.43, 4.94, 6.74, 9.19, 12.53\}~\unit{\giga\hertz}~\cite{Braun:2019gdo}.
There are four dominant sources of astrophysical foregrounds: galactic synchrotron radiation, galactic free-free radiation, spinning dust, and extragalactic point sources~\cite{Santos:2004ju}.
We simulate the first three foregrounds using the \textsc{pysm3} code~\cite{Thorne:2016ifb,Zonca:2021row,Pan-ExperimentGalacticScienceGroup:2025vcd} and the point sources using \textsc{epspy}~\cite{Mittal:2024mzv}. 
These foregrounds are simulated by beginning with template maps such as the \qty{408}{\mega\hertz} Haslam map from Ref.~\cite{Remazeilles:2014mba} and WMAP data~\cite{WMAP:2012fli}.
Then, small scale features are added to these templates under the standard assumption that $C_\ell$ obeys a power law at high-$\ell$ for each foreground. 
We assume that point sources brighter than $S_{\rm max} = \qty{0.1}{\milli\jansky}$ can be masked and are not a significant foreground; this choice is comparable to the threshold at the Low-Frequency Array (LOFAR)~\cite{Shimwell:2025tui}, which SKA should outperform.
Given the extremely small SKA beams~\cite{Braun:2019gdo}, we account for this masking by excluding these resolved sources from our mock foreground map~\cite{Wang:2005zj,Liu:2011hh}.  
However, a diffuse background of unresolved sources remains, which we model by using fits to their number density per unit flux, and their two-point correlation functions~\cite{Gervasi:2008rr, Hale:2023ust, Joseph:2019tti, Mittal:2024mzv}. 

Besides astrophysical foregrounds, we also add thermal noise to each map, which is drawn in each pixel from a Gaussian with $\sigma_{\rm rms}=\qty{7.5}{\micro\jansky\per\beam}$~\cite{Shimwell:2025tui}, which is greater than the expected thermal noise after $10$ hours of observing at SKA~\cite{Braun:2019gdo}.
We then mask the mock foreground maps, leaving only a \qty{20}{\degree} patch in the Southern Hemisphere unmasked, which greatly reduces the impact of galactic foregrounds.

Additionally, we also produce a mock map of $\delta_g$ that we cross-correlate with the $\gamma \to A'$ signal, constructed from the power spectrum of galaxies obtained from a halo occupation distribution (HOD) that is fit to the \textit{unWISE} galaxy survey~\cite{Kusiak:2022xkt}. 

\section{ILC Pipeline}
Now that we have modeled the mock null-hypothesis maps, we analyze them with the needlet ILC algorithm~\cite{Delabrouille:2008qd} that would be used to extract the $\gamma \to A'$ signal from the real SKA maps.
At radio frequencies, variations of this algorithm have also been proposed to search for the cosmological 21-cm signal \cite{Dai:2024bfa,Joseph:2024ush,DeCaro:2025qly}. 

The needlet ILC constructs the minimum-variance linear combination of our nine mock maps without reducing the strength of the dark photon signal of interest, which has frequency dependence $\propto \omega^{-1}$. 
We perform the needlet ILC algorithm using \textsc{pyilc}~\cite{McCarthy:2023hpa} (see~\citetalias{Baker:2025bnz} for more details).
The end product of this process is a single map, which is the optimal weighted sum of the input maps. 
We choose the normalization of the resulting map to be the magnitude of a signal-only map at $\omega_0 / (2\pi) =\qty{410}{\mega\hertz}$.
We run the ILC pipeline on our mock null-signal maps to obtain a post-ILC map, from which we compute the auto- and cross-power spectra (with the mock $\delta_g$ map), $C_\ell^{\mathrm{TT,obs}}$ and $C_\ell^{\mathrm{Tg,obs}}$, respectively.
These are shown in Fig.~\ref{fig:mccarthy_comparison} in orange, labeled ``fiducial'', and can be compared to several representative plots of $C_\ell^{\rm TX,pred}$ to get a sense of the sensitivity of SKA to dark photons.

Observe that $C_\ell^{\rm Tg}$ is negative, since $\gamma \to A'$ conversions lead to a deficit of photons, and therefore a negative correlation with the galaxy overdensity.
We note that the power spectra here differ from those in Refs.~\cite{Pirvu:2023lch,McCarthy:2024ozh} because of different definitions of the halo mass. 
In those works, the authors did not use a consistent definition of the halo mass when computing the halo mass function, bias, and concentration. 
Here, we use the virial mass definition in all computations and find that this change reduces the $C_\ell^{\rm Tg}$ by about 20\% relative to the results from Refs.~\cite{Pirvu:2023lch,McCarthy:2024ozh}. 
There is a smaller change to the autopower spectrum.

To validate our entire analysis procedure, we compare our results to previous work in the CMB frequency range. 
First, we apply our ILC pipeline to the \textsc{PR4 NPIPE} \textit{Planck} maps~\cite{Planck:2020olo} used in Refs.~\cite{McCarthy:2023cwg,McCarthy:2023hpa,McCarthy:2024ozh} to obtain a post-ILC $C^{\rm T T}_\ell$ angular power spectrum, shown in red in Fig.~\ref{fig:mccarthy_comparison}. 
This power spectrum agrees well with Fig.~1 of Ref.~\cite{McCarthy:2024ozh}, validating the ILC procedure.  
Next, we use the pipeline to produce mock maps at the \textit{Planck} frequencies and beam sizes.
These \textit{Planck} mock maps are then processed with our pipeline, and the resulting power spectrum is shown in blue in Fig.~\ref{fig:mccarthy_comparison}.
This mock power spectrum agrees relatively well with the actual \textit{Planck} power spectrum, demonstrating that our foreground modeling is able to accurately reproduce the \textit{Planck} sky. 

\section{Sensitivity of Radio Experiments}
\begin{figure}[t!]
    \centering
    \includegraphics[width=\linewidth]{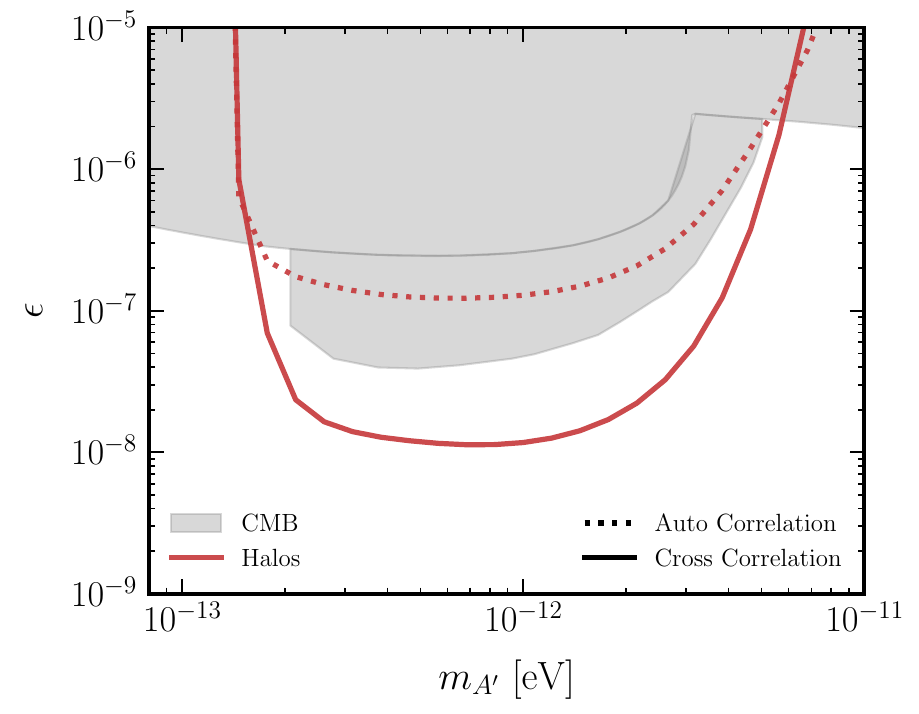}
    \caption{Projected sensitivity to $\gamma \to A'$ conversions in dark matter halos using the autocorrelation of SKA maps (dotted red) and in cross-correlation with the \textit{unWISE} galaxy catalog (solid red). Existing limits from CMB data are shown in gray.~\nblink{Limits}}
    \label{fig:ska_lims}
\end{figure}

To estimate the sensitivity of SKA to the dark photon signal, we construct a Gaussian likelihood of the form 
\begin{alignat}{1}
    -2 \ln \mathcal{L}_{\mathrm{TX}}(\epsilon) = \sum_{\ell,\ell'}\Delta C_\ell^{\rm TX} \cdot  \mathcal{C}_{\ell \ell'}^{-1}  \cdot \Delta C_{\ell'}^{\rm TX} \,,
\end{alignat}
where $\Delta C_\ell^{\rm TX} \equiv C_\ell^{\rm TX, pred} - C_\ell^{\rm TX, obs}$, for $\rm X \in \{T, g\}$. All $C_\ell$'s are binned with width $\Delta \ell = 150$. 
$\mathcal{C}_{\ell \ell'}$ is the Gaussian covariance between each $\ell$ bin, which we compute assuming Gaussian statistics with \textsc{pymaster} to account for mask effects~\cite{Alonso:2018jzx,Garcia-Garcia:2019bku,Nicola:2020lhi}.
We use these likelihoods to find the 95\% confidence upper limits for $\epsilon$, which are shown in Fig.~\ref{fig:ska_lims} for the full range of models and experiments that we consider in \citetalias{Baker:2025bnz}.

Overall, we find that radio experiments will have strong sensitivity to dark photon conversions for the models that we consider.
The sensitivity is greatest for the analysis that we focus on in this work, i.e.\ when considering the cross-correlation between $\gamma \to A'$ conversions in dark matter halos and low-redshift galaxies for $\qty{e-13}{\eV} \lesssim m_{A'}\lesssim \qty{5e-12}{\eV}$, shown as the red solid line in Fig.~\ref{fig:ska_lims}. 
SKA is sensitive to $\epsilon \sim 10^{-8}$ over this mass range, a factor of up to 4 improvement compared to the \textit{Planck} limit.

This improvement in sensitivity is due to two reasons. 
First, since the probability of conversion is proportional to $\omega^{-1}$, the signal is enhanced relative to CMB frequencies. 
Furthermore, since $\Delta T \propto x^{-1}(1-e^{-x})P_{\gamma \to A'}$, the observable signal at radio experiments does not suffer the exponential suppression that occurs at the high end of the CMB frequencies. 

Second, the smaller beams used in radio experiments drastically improve sensitivity. 
The autopower spectrum of the post-ILC maps constructed from \textit{Planck} data (Fig.~\ref{fig:mccarthy_comparison}, red) becomes large at high $\ell$, owing to the finite beam width of \textit{Planck}. 
At the lowest frequencies, the \textit{Planck} beams have a full-width half-maximum of $\theta_{\rm FWHM} \approx \qty{30}{\arcminute}$, leading to an exponential suppression of power in the maps at angular scales corresponding to $\ell \gtrsim 300$. 
In terms of the $\gamma \to A'$ signal, this means that the information used in the ILC procedure at large-$\ell$ comes from the high frequency maps; these have smaller beams, but also a weaker signal, leading to a loss in sensitivity and larger $C_\ell^{\rm TT, obs}$ at high-$\ell$. 

To confirm this interpretation, we compare our fiducial (Fig.~\ref{fig:mccarthy_comparison}, orange) post-ILC $C_\ell^{\rm TT}$ to the post-ILC $C_\ell^{\rm TT}$ that we would obtain at \textit{Planck} frequencies, but with a much smaller beam of $\theta_{\mathrm{FWHM}} = \qty{0.5}{\arcminute}$, comparable to the SKA beam size (Fig.~\ref{fig:mccarthy_comparison}, purple).
We find a significantly smaller post-ILC $C_\ell^{\rm TT}$ at high-$\ell$ with the smaller beam, demonstrating that the loss of sensitivity with \textit{Planck} at high-$\ell$ is due to the wide \textit{Planck} beams.

\section{Discussion}
In Fig.~\ref{fig:ska_lims} we plot the sensitivity of SKA and 21-cm global signal experiments to dark photon conversions in the IGM, which we discuss in detail in \citetalias{Baker:2025bnz}.
We first find that SKA cross-correlated with a high-redshift galaxy survey will be slightly more sensitive to dark photons than limits from CMB spectral distortions~\cite{Caputo:2020bdy}.
We compute this sensitivity by modeling $\gamma \to A'$ conversions in the IGM during the EoR and a catalog of high-redshift galaxies using the semianalytic code \textsc{21cmFAST}~\cite{Mesinger:2010ne, Park:2018ljd, Murray:2020trn}. 
We also estimate the sensitivity of SKA to $\gamma \to A'$ conversions that occur in the late Universe IGM, assuming that the baryon number density is distributed according to a lognormal distribution.
Although this is a crude model, we predict a strong $\gamma \to A'$ signal, motivating a more sophisticated hydrodynamical simulation study.

Beyond these imaging experiments, we further examine the sensitivity of 21-cm experiments to $\gamma \to A'$ conversions in \citetalias{Baker:2025bnz}. We find that a 21-cm global signal experiment is highly sensitive to $\gamma \to A'$ conversions if it can be disentangled from other foregrounds, but that the experimental details of a power spectrum experiment complicate the prospects for carrying out such a search there. 

We note that a study with actual data must consider several additional points.
In particular, Ref.~\cite{McCarthy:2024ozh} showed that the ILC can be slightly biased by foregrounds correlated with the $\gamma \to A'$ signal; such components must be explicitly deprojected.
We have not included this correction here, but based on the results of Ref.~\cite{McCarthy:2024ozh}, we expect only a modest weakening of our projected sensitivities.
Additionally, we have not accounted for a possible correlation between extragalactic radio emission and a galaxy survey. 
In a real SKA measurement, this correlation could produce a nonzero positive cross-correlation between the ILC map and a galaxy survey, which would artificially strengthen SKA's sensitivity.
Our results correspond to a successful removal of this bias, using, e.g. the subtraction procedure from Ref.~\cite{McCarthy:2024ozh}. Additionally, we perform a rudimentary estimate of this bias in \citetalias{Baker:2025bnz} and find that its magnitude is likely not appreciably larger than that found in Ref.~\cite{McCarthy:2024ozh}. In \citetalias{Baker:2025bnz}, we also study the impact of additional systematic uncertainties in the halo model and find that our results are conservative.

\section{Acknowledgments.} 
We would like to thank Colin Hill, Junwu Huang, Cristina Mondino and Julian Mu{\~n}oz for helpful conversations. The work in this paper makes extensive use of the \textsc{numpy}~\cite{Harris:2020xlr}, \textsc{scipy}~\cite{Virtanen:2019joe}, \textsc{matplotlib}~\cite{Hunter:2007ouj}, \textsc{astropy}~\cite{Astropy:2013muo,Astropy:2018wqo,Astropy:2022ucr}, \textsc{halomod}~\cite{Murray:2013qza,Murray:2020dcd}, \textsc{healpy}, and \textsc{healpix} packages~\cite{Gorski:2004by,Zonca:2019vzt}. We are also pleased to acknowledge that the computational work reported on in this paper was performed on the Shared Computing Cluster which is administered by Boston University's Research Computing Services.
E.B. was supported by the Boston University Dean's Fellowship program. E.B. and H.L. are supported by the U.S. Department of Energy under Grant No. DE-SC0026297 and the Cecile K. Dalton Career Development Professorship, endowed by Boston University trustee Nathaniel Dalton and Amy Gottlieb Dalton. 

\textbf{Data Availability.} --- 
The code and simulation data that support the findings of this article are available from Ref.~\cite{Baker_code_repo}.

\bibliography{sources}

\begin{thebibliography}{56}%
\makeatletter
\providecommand \@ifxundefined [1]{%
 \@ifx{#1\undefined}
}%
\providecommand \@ifnum [1]{%
 \ifnum #1\expandafter \@firstoftwo
 \else \expandafter \@secondoftwo
 \fi
}%
\providecommand \@ifx [1]{%
 \ifx #1\expandafter \@firstoftwo
 \else \expandafter \@secondoftwo
 \fi
}%
\providecommand \natexlab [1]{#1}%
\providecommand \enquote  [1]{``#1''}%
\providecommand \bibnamefont  [1]{#1}%
\providecommand \bibfnamefont [1]{#1}%
\providecommand \citenamefont [1]{#1}%
\providecommand \href@noop [0]{\@secondoftwo}%
\providecommand \href [0]{\begingroup \@sanitize@url \@href}%
\providecommand \@href[1]{\@@startlink{#1}\@@href}%
\providecommand \@@href[1]{\endgroup#1\@@endlink}%
\providecommand \@sanitize@url [0]{\catcode `\\12\catcode `\$12\catcode
  `\&12\catcode `\#12\catcode `\^12\catcode `\_12\catcode `\%12\relax}%
\providecommand \@@startlink[1]{}%
\providecommand \@@endlink[0]{}%
\providecommand \url  [0]{\begingroup\@sanitize@url \@url }%
\providecommand \@url [1]{\endgroup\@href {#1}{\urlprefix }}%
\providecommand \urlprefix  [0]{URL }%
\providecommand \Eprint [0]{\href }%
\providecommand \doibase [0]{https://doi.org/}%
\providecommand \selectlanguage [0]{\@gobble}%
\providecommand \bibinfo  [0]{\@secondoftwo}%
\providecommand \bibfield  [0]{\@secondoftwo}%
\providecommand \translation [1]{[#1]}%
\providecommand \BibitemOpen [0]{}%
\providecommand \bibitemStop [0]{}%
\providecommand \bibitemNoStop [0]{.\EOS\space}%
\providecommand \EOS [0]{\spacefactor3000\relax}%
\providecommand \BibitemShut  [1]{\csname bibitem#1\endcsname}%
\let\auto@bib@innerbib\@empty
\bibitem [{\citenamefont {Holdom}(1986)}]{Holdom:1985ag}%
  \BibitemOpen
  \bibfield  {author} {\bibinfo {author} {\bibfnamefont {B.}~\bibnamefont
  {Holdom}},\ }\bibfield  {title} {\bibinfo {title} {Two {{U}}(1)'s and
  {$\epsilon$} charge shifts},\ }\href
  {https://doi.org/10.1016/0370-2693(86)91377-8} {\bibfield  {journal}
  {\bibinfo  {journal} {Physics Letters B}\ }\textbf {\bibinfo {volume}
  {166}},\ \bibinfo {pages} {196} (\bibinfo {year} {1986})}\BibitemShut
  {NoStop}%
\bibitem [{\citenamefont {Mirizzi}\ \emph {et~al.}(2009)\citenamefont
  {Mirizzi}, \citenamefont {Redondo},\ and\ \citenamefont
  {Sigl}}]{Mirizzi:2009iz}%
  \BibitemOpen
  \bibfield  {author} {\bibinfo {author} {\bibfnamefont {A.}~\bibnamefont
  {Mirizzi}}, \bibinfo {author} {\bibfnamefont {J.}~\bibnamefont {Redondo}},\
  and\ \bibinfo {author} {\bibfnamefont {G.}~\bibnamefont {Sigl}},\ }\bibfield
  {title} {\bibinfo {title} {Microwave background constraints on mixing of
  photons with hidden photons},\ }\href
  {https://doi.org/10.1088/1475-7516/2009/03/026} {\bibfield  {journal}
  {\bibinfo  {journal} {Journal of Cosmology and Astroparticle Physics}\
  }\textbf {\bibinfo {volume} {2009}}\bibinfo  {number} { (03)},\ \bibinfo
  {pages} {026}}\BibitemShut {NoStop}%
\bibitem [{\citenamefont {Kunze}\ and\ \citenamefont
  {{Vazquez-Mozo}}(2015)}]{Kunze:2015noa}%
  \BibitemOpen
\bibfield  {number} {  }\bibfield  {author} {\bibinfo {author} {\bibfnamefont
  {K.~E.}\ \bibnamefont {Kunze}}\ and\ \bibinfo {author} {\bibfnamefont
  {M.~A.}\ \bibnamefont {{Vazquez-Mozo}}},\ }\bibfield  {title} {\bibinfo
  {title} {Constraints on hidden photons from current and future observations
  of {{CMB}} spectral distortions},\ }\href
  {https://doi.org/10.1088/1475-7516/2015/12/028} {\bibfield  {journal}
  {\bibinfo  {journal} {Journal of Cosmology and Astroparticle Physics}\
  }\textbf {\bibinfo {volume} {2015}}\bibfield  {number} {\bibinfo  {number} {
  (12)},\ \bibinfo {pages} {028}},\ }\Eprint {https://arxiv.org/abs/1507.02614}
  {arXiv:1507.02614 [astro-ph, physics:hep-ph]} \BibitemShut {NoStop}%
\bibitem [{\citenamefont {Caputo}\ \emph
  {et~al.}(2020{\natexlab{a}})\citenamefont {Caputo}, \citenamefont {Liu},
  \citenamefont {{Mishra-Sharma}},\ and\ \citenamefont
  {Ruderman}}]{Caputo:2020bdy}%
  \BibitemOpen
  \bibfield  {author} {\bibinfo {author} {\bibfnamefont {A.}~\bibnamefont
  {Caputo}}, \bibinfo {author} {\bibfnamefont {H.}~\bibnamefont {Liu}},
  \bibinfo {author} {\bibfnamefont {S.}~\bibnamefont {{Mishra-Sharma}}},\ and\
  \bibinfo {author} {\bibfnamefont {J.~T.}\ \bibnamefont {Ruderman}},\
  }\bibfield  {title} {\bibinfo {title} {Dark {{Photon Oscillations}} in {{Our
  Inhomogeneous Universe}}},\ }\href
  {https://doi.org/10.1103/PhysRevLett.125.221303} {\bibfield  {journal}
  {\bibinfo  {journal} {Physical Review Letters}\ }\textbf {\bibinfo {volume}
  {125}},\ \bibinfo {pages} {221303} (\bibinfo {year}
  {2020}{\natexlab{a}})}\BibitemShut {NoStop}%
\bibitem [{\citenamefont {Caputo}\ \emph
  {et~al.}(2020{\natexlab{b}})\citenamefont {Caputo}, \citenamefont {Liu},
  \citenamefont {{Mishra-Sharma}},\ and\ \citenamefont
  {Ruderman}}]{Caputo:2020rnx}%
  \BibitemOpen
  \bibfield  {author} {\bibinfo {author} {\bibfnamefont {A.}~\bibnamefont
  {Caputo}}, \bibinfo {author} {\bibfnamefont {H.}~\bibnamefont {Liu}},
  \bibinfo {author} {\bibfnamefont {S.}~\bibnamefont {{Mishra-Sharma}}},\ and\
  \bibinfo {author} {\bibfnamefont {J.~T.}\ \bibnamefont {Ruderman}},\
  }\bibfield  {title} {\bibinfo {title} {Modeling dark photon oscillations in
  our inhomogeneous {{Universe}}},\ }\href
  {https://doi.org/10.1103/PhysRevD.102.103533} {\bibfield  {journal} {\bibinfo
   {journal} {Physical Review D}\ }\textbf {\bibinfo {volume} {102}},\ \bibinfo
  {pages} {103533} (\bibinfo {year} {2020}{\natexlab{b}})}\BibitemShut
  {NoStop}%
\bibitem [{\citenamefont {Garcia}\ \emph {et~al.}(2020)\citenamefont {Garcia},
  \citenamefont {Bondarenko}, \citenamefont {Ploeckinger}, \citenamefont
  {Pradler},\ and\ \citenamefont {Sokolenko}}]{Garcia:2020qrp}%
  \BibitemOpen
  \bibfield  {author} {\bibinfo {author} {\bibfnamefont {A.~A.}\ \bibnamefont
  {Garcia}}, \bibinfo {author} {\bibfnamefont {K.}~\bibnamefont {Bondarenko}},
  \bibinfo {author} {\bibfnamefont {S.}~\bibnamefont {Ploeckinger}}, \bibinfo
  {author} {\bibfnamefont {J.}~\bibnamefont {Pradler}},\ and\ \bibinfo {author}
  {\bibfnamefont {A.}~\bibnamefont {Sokolenko}},\ }\bibfield  {title} {\bibinfo
  {title} {Effective photon mass and (dark) photon conversion in the
  inhomogeneous {{Universe}}},\ }\href
  {https://doi.org/10.1088/1475-7516/2020/10/011} {\bibfield  {journal}
  {\bibinfo  {journal} {JCAP}\ }\textbf {\bibinfo {volume} {10}},\ \bibinfo
  {pages} {011}}\BibitemShut {NoStop}%
\bibitem [{\citenamefont {Chluba}\ \emph {et~al.}(2024)\citenamefont {Chluba},
  \citenamefont {Cyr},\ and\ \citenamefont {Johnson}}]{Chluba:2024wui}%
  \BibitemOpen
  \bibfield  {author} {\bibinfo {author} {\bibfnamefont {J.}~\bibnamefont
  {Chluba}}, \bibinfo {author} {\bibfnamefont {B.}~\bibnamefont {Cyr}},\ and\
  \bibinfo {author} {\bibfnamefont {M.~C.}\ \bibnamefont {Johnson}},\
  }\bibfield  {title} {\bibinfo {title} {Revisiting dark photon constraints
  from {{CMB}} spectral distortions},\ }\href
  {https://doi.org/10.1093/mnras/stae2464} {\bibfield  {journal} {\bibinfo
  {journal} {Mon. Not. Roy. Astron. Soc.}\ }\textbf {\bibinfo {volume} {535}},\
  \bibinfo {pages} {1874} (\bibinfo {year} {2024})}\BibitemShut {NoStop}%
\bibitem [{\citenamefont {Arsenadze}\ \emph {et~al.}(2025)\citenamefont
  {Arsenadze}, \citenamefont {Caputo}, \citenamefont {Gan}, \citenamefont
  {Liu},\ and\ \citenamefont {Ruderman}}]{Arsenadze:2024ywr}%
  \BibitemOpen
  \bibfield  {author} {\bibinfo {author} {\bibfnamefont {G.}~\bibnamefont
  {Arsenadze}}, \bibinfo {author} {\bibfnamefont {A.}~\bibnamefont {Caputo}},
  \bibinfo {author} {\bibfnamefont {X.}~\bibnamefont {Gan}}, \bibinfo {author}
  {\bibfnamefont {H.}~\bibnamefont {Liu}},\ and\ \bibinfo {author}
  {\bibfnamefont {J.~T.}\ \bibnamefont {Ruderman}},\ }\bibfield  {title}
  {\bibinfo {title} {Shaping dark photon spectral distortions},\ }\href
  {https://doi.org/10.1007/JHEP03(2025)018} {\bibfield  {journal} {\bibinfo
  {journal} {JHEP}\ }\textbf {\bibinfo {volume} {03}},\ \bibinfo {pages}
  {018}}\BibitemShut {NoStop}%
\bibitem [{\citenamefont {P{\^i}rvu}\ \emph {et~al.}(2024)\citenamefont
  {P{\^i}rvu}, \citenamefont {Huang},\ and\ \citenamefont
  {Johnson}}]{Pirvu:2023lch}%
  \BibitemOpen
  \bibfield  {author} {\bibinfo {author} {\bibfnamefont {D.}~\bibnamefont
  {P{\^i}rvu}}, \bibinfo {author} {\bibfnamefont {J.}~\bibnamefont {Huang}},\
  and\ \bibinfo {author} {\bibfnamefont {M.~C.}\ \bibnamefont {Johnson}},\
  }\bibfield  {title} {\bibinfo {title} {Patchy screening of the {{CMB}} from
  dark photons},\ }\href {https://doi.org/10.1088/1475-7516/2024/01/019}
  {\bibfield  {journal} {\bibinfo  {journal} {Journal of Cosmology and
  Astroparticle Physics}\ }\textbf {\bibinfo {volume} {2024}}\bibinfo  {number}
  { (01)},\ \bibinfo {pages} {019}}\BibitemShut {NoStop}%
\bibitem [{\citenamefont {McCarthy}\ \emph {et~al.}(2024)\citenamefont
  {McCarthy}, \citenamefont {Pirvu}, \citenamefont {Hill}, \citenamefont
  {Huang}, \citenamefont {Johnson},\ and\ \citenamefont
  {Rogers}}]{McCarthy:2024ozh}%
  \BibitemOpen
\bibfield  {number} {  }\bibfield  {author} {\bibinfo {author} {\bibfnamefont
  {F.}~\bibnamefont {McCarthy}}, \bibinfo {author} {\bibfnamefont
  {D.}~\bibnamefont {Pirvu}}, \bibinfo {author} {\bibfnamefont {J.~C.}\
  \bibnamefont {Hill}}, \bibinfo {author} {\bibfnamefont {J.}~\bibnamefont
  {Huang}}, \bibinfo {author} {\bibfnamefont {M.~C.}\ \bibnamefont {Johnson}},\
  and\ \bibinfo {author} {\bibfnamefont {K.~K.}\ \bibnamefont {Rogers}},\
  }\bibfield  {title} {\bibinfo {title} {Dark {{Photon Limits}} from {{Patchy
  Dark Screening}} of the {{Cosmic Microwave Background}}},\ }\href
  {https://doi.org/10.1103/PhysRevLett.133.141003} {\bibfield  {journal}
  {\bibinfo  {journal} {Phys. Rev. Lett.}\ }\textbf {\bibinfo {volume} {133}},\
  \bibinfo {pages} {141003} (\bibinfo {year} {2024})}\BibitemShut {NoStop}%
\bibitem [{\citenamefont {{Aramburo-Garcia}}\ \emph {et~al.}(2024)\citenamefont
  {{Aramburo-Garcia}}, \citenamefont {Bondarenko}, \citenamefont {Boyarsky},
  \citenamefont {Kashko}, \citenamefont {Pradler}, \citenamefont {Sokolenko},
  \citenamefont {Kugel}, \citenamefont {Schaller},\ and\ \citenamefont
  {Schaye}}]{Aramburo-Garcia:2024cbz}%
  \BibitemOpen
  \bibfield  {author} {\bibinfo {author} {\bibfnamefont {A.}~\bibnamefont
  {{Aramburo-Garcia}}}, \bibinfo {author} {\bibfnamefont {K.}~\bibnamefont
  {Bondarenko}}, \bibinfo {author} {\bibfnamefont {A.}~\bibnamefont
  {Boyarsky}}, \bibinfo {author} {\bibfnamefont {P.}~\bibnamefont {Kashko}},
  \bibinfo {author} {\bibfnamefont {J.}~\bibnamefont {Pradler}}, \bibinfo
  {author} {\bibfnamefont {A.}~\bibnamefont {Sokolenko}}, \bibinfo {author}
  {\bibfnamefont {R.}~\bibnamefont {Kugel}}, \bibinfo {author} {\bibfnamefont
  {M.}~\bibnamefont {Schaller}},\ and\ \bibinfo {author} {\bibfnamefont
  {J.}~\bibnamefont {Schaye}},\ }\bibfield  {title} {\bibinfo {title} {Dark
  photon constraints from {{CMB}} temperature anisotropies},\ }\href
  {https://doi.org/10.1088/1475-7516/2024/11/049} {\bibfield  {journal}
  {\bibinfo  {journal} {JCAP}\ }\textbf {\bibinfo {volume} {11}},\ \bibinfo
  {pages} {049}}\BibitemShut {NoStop}%
\bibitem [{\citenamefont {Baker}\ and\ \citenamefont
  {Liu}(2026{\natexlab{a}})}]{Baker:2025bnz}%
  \BibitemOpen
  \bibfield  {author} {\bibinfo {author} {\bibfnamefont {E.}~\bibnamefont
  {Baker}}\ and\ \bibinfo {author} {\bibfnamefont {H.}~\bibnamefont {Liu}},\
  }\bibfield  {title} {\bibinfo {title} {Dark photons in the radio sky. {{II}}.
  {{Resonant}} conversions in the intergalactic medium},\ }\href
  {https://doi.org/10.1103/q492-rszq} {\bibfield  {journal} {\bibinfo
  {journal} {Phys. Rev. D}\ }\textbf {\bibinfo {volume} {114}},\ \bibinfo
  {pages} {043014} (\bibinfo {year} {2026}{\natexlab{a}})},\ \Eprint
  {https://arxiv.org/abs/2511.09637} {arXiv:2511.09637 [astro-ph.CO]}
  \BibitemShut {NoStop}%
\bibitem [{\citenamefont {Thorne}\ \emph {et~al.}(2017)\citenamefont {Thorne},
  \citenamefont {Dunkley}, \citenamefont {Alonso},\ and\ \citenamefont
  {Naess}}]{Thorne:2016ifb}%
  \BibitemOpen
  \bibfield  {author} {\bibinfo {author} {\bibfnamefont {B.}~\bibnamefont
  {Thorne}}, \bibinfo {author} {\bibfnamefont {J.}~\bibnamefont {Dunkley}},
  \bibinfo {author} {\bibfnamefont {D.}~\bibnamefont {Alonso}},\ and\ \bibinfo
  {author} {\bibfnamefont {S.}~\bibnamefont {Naess}},\ }\bibfield  {title}
  {\bibinfo {title} {The {{Python Sky Model}}: Software for simulating the
  {{Galactic}} microwave sky},\ }\href {https://doi.org/10.1093/mnras/stx949}
  {\bibfield  {journal} {\bibinfo  {journal} {Monthly Notices of the Royal
  Astronomical Society}\ }\textbf {\bibinfo {volume} {469}},\ \bibinfo {pages}
  {2821} (\bibinfo {year} {2017})},\ \Eprint {https://arxiv.org/abs/1608.02841}
  {arXiv:1608.02841 [astro-ph]} \BibitemShut {NoStop}%
\bibitem [{\citenamefont {Zonca}\ \emph {et~al.}(2021)\citenamefont {Zonca},
  \citenamefont {Thorne}, \citenamefont {Krachmalnicoff},\ and\ \citenamefont
  {Borrill}}]{Zonca:2021row}%
  \BibitemOpen
  \bibfield  {author} {\bibinfo {author} {\bibfnamefont {A.}~\bibnamefont
  {Zonca}}, \bibinfo {author} {\bibfnamefont {B.}~\bibnamefont {Thorne}},
  \bibinfo {author} {\bibfnamefont {N.}~\bibnamefont {Krachmalnicoff}},\ and\
  \bibinfo {author} {\bibfnamefont {J.}~\bibnamefont {Borrill}},\ }\bibfield
  {title} {\bibinfo {title} {The {{Python Sky Model}} 3 software},\ }\href
  {https://doi.org/10.21105/joss.03783} {\bibfield  {journal} {\bibinfo
  {journal} {Journal of Open Source Software}\ }\textbf {\bibinfo {volume}
  {6}},\ \bibinfo {pages} {3783} (\bibinfo {year} {2021})},\ \Eprint
  {https://arxiv.org/abs/2108.01444} {arXiv:2108.01444 [astro-ph]} \BibitemShut
  {NoStop}%
\bibitem [{\citenamefont {Borrill}\ \emph {et~al.}(2026)\citenamefont {Borrill}
  \emph {et~al.}}]{Pan-ExperimentGalacticScienceGroup:2025vcd}%
  \BibitemOpen
  \bibfield  {author} {\bibinfo {author} {\bibfnamefont {J.}~\bibnamefont
  {Borrill}} \emph {et~al.},\ }\bibfield  {title} {\bibinfo {title} {Full-sky
  {{Models}} of {{Galactic Microwave Emission}} and {{Polarization}} at
  {{Sub-arcminute Scales}} for the {{Python Sky Model}}},\ }\href
  {https://doi.org/10.3847/1538-4357/adf212} {\bibfield  {journal} {\bibinfo
  {journal} {Astrophys. J.}\ }\textbf {\bibinfo {volume} {991}},\ \bibinfo
  {pages} {23} (\bibinfo {year} {2026})}\BibitemShut {NoStop}%
\bibitem [{\citenamefont {Gervasi}\ \emph {et~al.}(2008)\citenamefont
  {Gervasi}, \citenamefont {Tartari}, \citenamefont {Zannoni}, \citenamefont
  {Boella},\ and\ \citenamefont {Sironi}}]{Gervasi:2008rr}%
  \BibitemOpen
  \bibfield  {author} {\bibinfo {author} {\bibfnamefont {M.}~\bibnamefont
  {Gervasi}}, \bibinfo {author} {\bibfnamefont {A.}~\bibnamefont {Tartari}},
  \bibinfo {author} {\bibfnamefont {M.}~\bibnamefont {Zannoni}}, \bibinfo
  {author} {\bibfnamefont {G.}~\bibnamefont {Boella}},\ and\ \bibinfo {author}
  {\bibfnamefont {G.}~\bibnamefont {Sironi}},\ }\bibfield  {title} {\bibinfo
  {title} {The contribution of the {{Unresolved Extragalactic Radio Sources}}
  to the {{Brightness Temperature}} of the sky},\ }\href
  {https://doi.org/10.1086/588628} {\bibfield  {journal} {\bibinfo  {journal}
  {Astrophys. J.}\ }\textbf {\bibinfo {volume} {682}},\ \bibinfo {pages} {223}
  (\bibinfo {year} {2008})}\BibitemShut {NoStop}%
\bibitem [{\citenamefont {Mittal}\ \emph {et~al.}(2024)\citenamefont {Mittal},
  \citenamefont {Kulkarni}, \citenamefont {Anstey},\ and\ \citenamefont
  {{de~Lera~Acedo}}}]{Mittal:2024mzv}%
  \BibitemOpen
  \bibfield  {author} {\bibinfo {author} {\bibfnamefont {S.}~\bibnamefont
  {Mittal}}, \bibinfo {author} {\bibfnamefont {G.}~\bibnamefont {Kulkarni}},
  \bibinfo {author} {\bibfnamefont {D.}~\bibnamefont {Anstey}},\ and\ \bibinfo
  {author} {\bibfnamefont {E.}~\bibnamefont {{de~Lera~Acedo}}},\ }\bibfield
  {title} {\bibinfo {title} {Impact of extragalactic point sources on the
  low-frequency sky spectrum and cosmic dawn global 21-cm measurements},\
  }\href {https://doi.org/10.1093/mnras/stae2111} {\bibfield  {journal}
  {\bibinfo  {journal} {Monthly Notices of the Royal Astronomical Society}\
  }\textbf {\bibinfo {volume} {534}},\ \bibinfo {pages} {1317} (\bibinfo {year}
  {2024})}\BibitemShut {NoStop}%
\bibitem [{\citenamefont {Bennett}\ \emph {et~al.}(2003)\citenamefont {Bennett}
  \emph {et~al.}}]{WMAP:2003cmr}%
  \BibitemOpen
  \bibfield  {author} {\bibinfo {author} {\bibfnamefont {C.}~\bibnamefont
  {Bennett}} \emph {et~al.} (\bibinfo {collaboration} {WMAP Collaboration}),\
  }\bibfield  {title} {\bibinfo {title} {First year {{Wilkinson Microwave
  Anisotropy Probe}} ({{WMAP}}) observations: {{Foreground}} emission},\ }\href
  {https://doi.org/10.1086/377252} {\bibfield  {journal} {\bibinfo  {journal}
  {Astrophys. J. Suppl.}\ }\textbf {\bibinfo {volume} {148}},\ \bibinfo {pages}
  {97} (\bibinfo {year} {2003})}\BibitemShut {NoStop}%
\bibitem [{\citenamefont {Tegmark}\ \emph {et~al.}(2003)\citenamefont
  {Tegmark}, \citenamefont {de~{Oliveira-Costa}},\ and\ \citenamefont
  {Hamilton}}]{Tegmark:2003ve}%
  \BibitemOpen
  \bibfield  {author} {\bibinfo {author} {\bibfnamefont {M.}~\bibnamefont
  {Tegmark}}, \bibinfo {author} {\bibfnamefont {A.}~\bibnamefont
  {de~{Oliveira-Costa}}},\ and\ \bibinfo {author} {\bibfnamefont
  {A.}~\bibnamefont {Hamilton}},\ }\bibfield  {title} {\bibinfo {title} {A high
  resolution foreground cleaned {{CMB}} map from {{WMAP}}},\ }\href
  {https://doi.org/10.1103/PhysRevD.68.123523} {\bibfield  {journal} {\bibinfo
  {journal} {Physical Review D}\ }\textbf {\bibinfo {volume} {68}},\ \bibinfo
  {pages} {123523} (\bibinfo {year} {2003})},\ \Eprint
  {https://arxiv.org/abs/astro-ph/0302496} {arXiv:astro-ph/0302496}
  \BibitemShut {NoStop}%
\bibitem [{\citenamefont {Delabrouille}\ \emph {et~al.}(2009)\citenamefont
  {Delabrouille}, \citenamefont {Cardoso}, \citenamefont {Le~Jeune},
  \citenamefont {Betoule}, \citenamefont {Fay},\ and\ \citenamefont
  {Guilloux}}]{Delabrouille:2008qd}%
  \BibitemOpen
  \bibfield  {author} {\bibinfo {author} {\bibfnamefont {J.}~\bibnamefont
  {Delabrouille}}, \bibinfo {author} {\bibfnamefont {J.~F.}\ \bibnamefont
  {Cardoso}}, \bibinfo {author} {\bibfnamefont {M.}~\bibnamefont {Le~Jeune}},
  \bibinfo {author} {\bibfnamefont {M.}~\bibnamefont {Betoule}}, \bibinfo
  {author} {\bibfnamefont {G.}~\bibnamefont {Fay}},\ and\ \bibinfo {author}
  {\bibfnamefont {F.}~\bibnamefont {Guilloux}},\ }\bibfield  {title} {\bibinfo
  {title} {A full sky, low foreground, high resolution {{CMB}} map from
  {{WMAP}}},\ }\href {https://doi.org/10.1051/0004-6361:200810514} {\bibfield
  {journal} {\bibinfo  {journal} {Astronomy and Astrophysics}\ }\textbf
  {\bibinfo {volume} {493}},\ \bibinfo {pages} {835} (\bibinfo {year}
  {2009})}\BibitemShut {NoStop}%
\bibitem [{\citenamefont {McCarthy}\ and\ \citenamefont
  {Hill}(2024)}]{McCarthy:2023hpa}%
  \BibitemOpen
  \bibfield  {author} {\bibinfo {author} {\bibfnamefont {F.}~\bibnamefont
  {McCarthy}}\ and\ \bibinfo {author} {\bibfnamefont {J.~C.}\ \bibnamefont
  {Hill}},\ }\bibfield  {title} {\bibinfo {title} {Component-separated,
  {{CIB-cleaned}} thermal {{Sunyaev-Zel}}'dovich maps from {{{\emph{Planck}}}}
  {{PR4}} data with a flexible public needlet {{ILC}} pipeline},\ }\href
  {https://doi.org/10.1103/PhysRevD.109.023528} {\bibfield  {journal} {\bibinfo
   {journal} {Physical Review D}\ }\textbf {\bibinfo {volume} {109}},\ \bibinfo
  {pages} {023528} (\bibinfo {year} {2024})}\BibitemShut {NoStop}%
\bibitem [{\citenamefont {Fixsen}(2009)}]{Fixsen:2009ug}%
  \BibitemOpen
  \bibfield  {author} {\bibinfo {author} {\bibfnamefont {D.~J.}\ \bibnamefont
  {Fixsen}},\ }\bibfield  {title} {\bibinfo {title} {The {{Temperature}} of the
  {{Cosmic Microwave Background}}},\ }\href
  {https://doi.org/10.1088/0004-637X/707/2/916} {\bibfield  {journal} {\bibinfo
   {journal} {The Astrophysical Journal}\ }\textbf {\bibinfo {volume} {707}},\
  \bibinfo {pages} {916} (\bibinfo {year} {2009})},\ \Eprint
  {https://arxiv.org/abs/0911.1955} {arXiv:0911.1955 [astro-ph]} \BibitemShut
  {NoStop}%
\bibitem [{\citenamefont {Battaglia}(2016)}]{Battaglia:2016xbi}%
  \BibitemOpen
  \bibfield  {author} {\bibinfo {author} {\bibfnamefont {N.}~\bibnamefont
  {Battaglia}},\ }\bibfield  {title} {\bibinfo {title} {The {{Tau}} of {{Galaxy
  Clusters}}},\ }\href {https://doi.org/10.1088/1475-7516/2016/08/058}
  {\bibfield  {journal} {\bibinfo  {journal} {Journal of Cosmology and
  Astroparticle Physics}\ }\textbf {\bibinfo {volume} {2016}}\bibfield
  {number} {\bibinfo  {number} { (08)},\ \bibinfo {pages} {058}},\ }\Eprint
  {https://arxiv.org/abs/1607.02442} {arXiv:1607.02442 [astro-ph]} \BibitemShut
  {NoStop}%
\bibitem [{\citenamefont {Kusiak}\ \emph {et~al.}(2022)\citenamefont {Kusiak},
  \citenamefont {Bolliet}, \citenamefont {Krolewski},\ and\ \citenamefont
  {Hill}}]{Kusiak:2022xkt}%
  \BibitemOpen
  \bibfield  {author} {\bibinfo {author} {\bibfnamefont {A.}~\bibnamefont
  {Kusiak}}, \bibinfo {author} {\bibfnamefont {B.}~\bibnamefont {Bolliet}},
  \bibinfo {author} {\bibfnamefont {A.}~\bibnamefont {Krolewski}},\ and\
  \bibinfo {author} {\bibfnamefont {J.~C.}\ \bibnamefont {Hill}},\ }\bibfield
  {title} {\bibinfo {title} {Constraining the galaxy-halo connection of
  infrared-selected {{unWISE}} galaxies with galaxy clustering and
  galaxy-{{CMB}} lensing power spectra},\ }\href
  {https://doi.org/10.1103/PhysRevD.106.123517} {\bibfield  {journal} {\bibinfo
   {journal} {Physical Review D}\ }\textbf {\bibinfo {volume} {106}},\ \bibinfo
  {pages} {123517} (\bibinfo {year} {2022})},\ \Eprint
  {https://arxiv.org/abs/2203.12583} {arXiv:2203.12583 [astro-ph]} \BibitemShut
  {NoStop}%
\bibitem [{hea()}]{healpix}%
  \BibitemOpen
  \href@noop {} {}\bibinfo {howpublished}
  {\url{https://healpix.sourceforge.io/}}\BibitemShut {NoStop}%
\bibitem [{\citenamefont {Braun}\ \emph {et~al.}(2019)\citenamefont {Braun},
  \citenamefont {Bonaldi}, \citenamefont {Bourke}, \citenamefont {Keane},\ and\
  \citenamefont {Wagg}}]{Braun:2019gdo}%
  \BibitemOpen
  \bibfield  {author} {\bibinfo {author} {\bibfnamefont {R.}~\bibnamefont
  {Braun}}, \bibinfo {author} {\bibfnamefont {A.}~\bibnamefont {Bonaldi}},
  \bibinfo {author} {\bibfnamefont {T.}~\bibnamefont {Bourke}}, \bibinfo
  {author} {\bibfnamefont {E.}~\bibnamefont {Keane}},\ and\ \bibinfo {author}
  {\bibfnamefont {J.}~\bibnamefont {Wagg}},\ }\href
  {https://doi.org/10.48550/arXiv.1912.12699} {\bibinfo {title} {Anticipated
  {{Performance}} of the {{Square Kilometre Array}} -- {{Phase}} 1 ({{SKA1}})}}
  (\bibinfo {year} {2019}),\ \Eprint {https://arxiv.org/abs/1912.12699}
  {arXiv:1912.12699 [astro-ph.IM]} \BibitemShut {NoStop}%
\bibitem [{\citenamefont {Santos}\ \emph {et~al.}(2005)\citenamefont {Santos},
  \citenamefont {Cooray},\ and\ \citenamefont {Knox}}]{Santos:2004ju}%
  \BibitemOpen
  \bibfield  {author} {\bibinfo {author} {\bibfnamefont {M.~G.}\ \bibnamefont
  {Santos}}, \bibinfo {author} {\bibfnamefont {A.}~\bibnamefont {Cooray}},\
  and\ \bibinfo {author} {\bibfnamefont {L.}~\bibnamefont {Knox}},\ }\bibfield
  {title} {\bibinfo {title} {Multifrequency analysis of 21 cm fluctuations from
  the era of reionization},\ }\href {https://doi.org/10.1086/429857} {\bibfield
   {journal} {\bibinfo  {journal} {Astrophys. J.}\ }\textbf {\bibinfo {volume}
  {625}},\ \bibinfo {pages} {575} (\bibinfo {year} {2005})}\BibitemShut
  {NoStop}%
\bibitem [{\citenamefont {Remazeilles}\ \emph {et~al.}(2015)\citenamefont
  {Remazeilles}, \citenamefont {Dickinson}, \citenamefont {Banday},
  \citenamefont {{Bigot-Sazy}},\ and\ \citenamefont
  {Ghosh}}]{Remazeilles:2014mba}%
  \BibitemOpen
  \bibfield  {author} {\bibinfo {author} {\bibfnamefont {M.}~\bibnamefont
  {Remazeilles}}, \bibinfo {author} {\bibfnamefont {C.}~\bibnamefont
  {Dickinson}}, \bibinfo {author} {\bibfnamefont {A.~J.}\ \bibnamefont
  {Banday}}, \bibinfo {author} {\bibfnamefont {M.-A.}\ \bibnamefont
  {{Bigot-Sazy}}},\ and\ \bibinfo {author} {\bibfnamefont {T.}~\bibnamefont
  {Ghosh}},\ }\bibfield  {title} {\bibinfo {title} {An improved
  source-subtracted and destriped 408-{{MHz}} all-sky map},\ }\href
  {https://doi.org/10.1093/mnras/stv1274} {\bibfield  {journal} {\bibinfo
  {journal} {Monthly Notices of the Royal Astronomical Society}\ }\textbf
  {\bibinfo {volume} {451}},\ \bibinfo {pages} {4311} (\bibinfo {year}
  {2015})}\BibitemShut {NoStop}%
\bibitem [{\citenamefont {Bennett}\ \emph {et~al.}(2013)\citenamefont {Bennett}
  \emph {et~al.}}]{WMAP:2012fli}%
  \BibitemOpen
  \bibfield  {author} {\bibinfo {author} {\bibfnamefont {C.~L.}\ \bibnamefont
  {Bennett}} \emph {et~al.} (\bibinfo {collaboration} {WMAP Collaboration}),\
  }\bibfield  {title} {\bibinfo {title} {Nine-year {{Wilkinson Microwave
  Anisotropy Probe}} ({{WMAP}}) {{Observations}}: {{Final Maps}} and
  {{Results}}},\ }\href {https://doi.org/10.1088/0067-0049/208/2/20} {\bibfield
   {journal} {\bibinfo  {journal} {The Astrophysical Journal Supplement
  Series}\ }\textbf {\bibinfo {volume} {208}},\ \bibinfo {pages} {20} (\bibinfo
  {year} {2013})}\BibitemShut {NoStop}%
\bibitem [{\citenamefont {Shimwell}\ \emph {et~al.}(2025)\citenamefont
  {Shimwell} \emph {et~al.}}]{Shimwell:2025tui}%
  \BibitemOpen
  \bibfield  {author} {\bibinfo {author} {\bibfnamefont {T.~W.}\ \bibnamefont
  {Shimwell}} \emph {et~al.},\ }\bibfield  {title} {\bibinfo {title} {The
  {{LOFAR Two-metre Sky Survey}}: {{Deep Fields Data Release}} 2: {{I}}. {{The
  ELAIS-N1}} field},\ }\href {https://doi.org/10.1051/0004-6361/202452930}
  {\bibfield  {journal} {\bibinfo  {journal} {Astronomy \& Astrophysics}\
  }\textbf {\bibinfo {volume} {695}},\ \bibinfo {pages} {A80} (\bibinfo {year}
  {2025})}\BibitemShut {NoStop}%
\bibitem [{\citenamefont {Wang}\ \emph {et~al.}(2006)\citenamefont {Wang},
  \citenamefont {Tegmark}, \citenamefont {Santos},\ and\ \citenamefont
  {Knox}}]{Wang:2005zj}%
  \BibitemOpen
  \bibfield  {author} {\bibinfo {author} {\bibfnamefont {X.}~\bibnamefont
  {Wang}}, \bibinfo {author} {\bibfnamefont {M.}~\bibnamefont {Tegmark}},
  \bibinfo {author} {\bibfnamefont {M.~G.}\ \bibnamefont {Santos}},\ and\
  \bibinfo {author} {\bibfnamefont {L.}~\bibnamefont {Knox}},\ }\bibfield
  {title} {\bibinfo {title} {21 cm {{Tomography}} with {{Foregrounds}}},\
  }\href {https://doi.org/10.1086/506597} {\bibfield  {journal} {\bibinfo
  {journal} {The Astrophysical Journal}\ }\textbf {\bibinfo {volume} {650}},\
  \bibinfo {pages} {529} (\bibinfo {year} {2006})}\BibitemShut {NoStop}%
\bibitem [{\citenamefont {Liu}\ and\ \citenamefont
  {Tegmark}(2011)}]{Liu:2011hh}%
  \BibitemOpen
  \bibfield  {author} {\bibinfo {author} {\bibfnamefont {A.}~\bibnamefont
  {Liu}}\ and\ \bibinfo {author} {\bibfnamefont {M.}~\bibnamefont {Tegmark}},\
  }\bibfield  {title} {\bibinfo {title} {A method for 21 cm power spectrum
  estimation in the presence of foregrounds},\ }\href
  {https://doi.org/10.1103/PhysRevD.83.103006} {\bibfield  {journal} {\bibinfo
  {journal} {Physical Review D}\ }\textbf {\bibinfo {volume} {83}},\ \bibinfo
  {pages} {103006} (\bibinfo {year} {2011})}\BibitemShut {NoStop}%
\bibitem [{\citenamefont {Hale}\ \emph {et~al.}(2024)\citenamefont {Hale} \emph
  {et~al.}}]{Hale:2023ust}%
  \BibitemOpen
  \bibfield  {author} {\bibinfo {author} {\bibfnamefont {C.~L.}\ \bibnamefont
  {Hale}} \emph {et~al.},\ }\bibfield  {title} {\bibinfo {title} {Cosmology
  from {{LOFAR Two-metre Sky Survey Data Release}} 2: Angular clustering of
  radio sources},\ }\href {https://doi.org/10.1093/mnras/stad3088} {\bibfield
  {journal} {\bibinfo  {journal} {Monthly Notices of the Royal Astronomical
  Society}\ }\textbf {\bibinfo {volume} {527}},\ \bibinfo {pages} {6540}
  (\bibinfo {year} {2024})}\BibitemShut {NoStop}%
\bibitem [{\citenamefont {Joseph}\ \emph {et~al.}(2020)\citenamefont {Joseph},
  \citenamefont {Trott}, \citenamefont {Wayth},\ and\ \citenamefont
  {Nasirudin}}]{Joseph:2019tti}%
  \BibitemOpen
  \bibfield  {author} {\bibinfo {author} {\bibfnamefont {R.~C.}\ \bibnamefont
  {Joseph}}, \bibinfo {author} {\bibfnamefont {C.~M.}\ \bibnamefont {Trott}},
  \bibinfo {author} {\bibfnamefont {R.~B.}\ \bibnamefont {Wayth}},\ and\
  \bibinfo {author} {\bibfnamefont {A.}~\bibnamefont {Nasirudin}},\ }\bibfield
  {title} {\bibinfo {title} {Calibration and 21-cm {{Power Spectrum
  Estimation}} in the {{Presence}} of {{Antenna Beam Variations}}},\ }\href
  {https://doi.org/10.1093/mnras/stz3375} {\bibfield  {journal} {\bibinfo
  {journal} {Monthly Notices of the Royal Astronomical Society}\ }\textbf
  {\bibinfo {volume} {492}},\ \bibinfo {pages} {2017} (\bibinfo {year}
  {2020})},\ \Eprint {https://arxiv.org/abs/1911.13088} {arXiv:1911.13088
  [astro-ph]} \BibitemShut {NoStop}%
\bibitem [{\citenamefont {Dai}\ and\ \citenamefont {Ma}(2025)}]{Dai:2024bfa}%
  \BibitemOpen
  \bibfield  {author} {\bibinfo {author} {\bibfnamefont {W.-M.}\ \bibnamefont
  {Dai}}\ and\ \bibinfo {author} {\bibfnamefont {Y.-Z.}\ \bibnamefont {Ma}},\
  }\bibfield  {title} {\bibinfo {title} {Expanded {{Generalized Needlet
  Internal Linear Combination}} ({{eGNILC}}) {{Framework}} for the 21-cm
  {{Foreground Removal}}},\ }\href {https://doi.org/10.3847/1538-4365/ad9604}
  {\bibfield  {journal} {\bibinfo  {journal} {The Astrophysical Journal
  Supplement Series}\ }\textbf {\bibinfo {volume} {276}},\ \bibinfo {pages}
  {33} (\bibinfo {year} {2025})},\ \Eprint {https://arxiv.org/abs/2411.16899}
  {arXiv:2411.16899 [astro-ph]} \BibitemShut {NoStop}%
\bibitem [{\citenamefont {Joseph}\ and\ \citenamefont
  {Saha}(2025)}]{Joseph:2024ush}%
  \BibitemOpen
  \bibfield  {author} {\bibinfo {author} {\bibfnamefont {A.}~\bibnamefont
  {Joseph}}\ and\ \bibinfo {author} {\bibfnamefont {R.}~\bibnamefont {Saha}},\
  }\bibfield  {title} {\bibinfo {title} {Foreground removal and angular power
  spectrum estimation of 21 cm signal using harmonic space {{ILC}} method},\
  }\href {https://doi.org/10.3847/1538-4357/adb3a9} {\bibfield  {journal}
  {\bibinfo  {journal} {The Astrophysical Journal}\ }\textbf {\bibinfo {volume}
  {982}},\ \bibinfo {pages} {49} (\bibinfo {year} {2025})},\ \Eprint
  {https://arxiv.org/abs/2405.02806} {arXiv:2405.02806 [astro-ph]} \BibitemShut
  {NoStop}%
\bibitem [{\citenamefont {Caro}\ \emph {et~al.}(2025)\citenamefont {Caro},
  \citenamefont {Carucci}, \citenamefont {Camera}, \citenamefont
  {Remazeilles},\ and\ \citenamefont {Carbone}}]{DeCaro:2025qly}%
  \BibitemOpen
  \bibfield  {author} {\bibinfo {author} {\bibfnamefont {B.~D.}\ \bibnamefont
  {Caro}}, \bibinfo {author} {\bibfnamefont {I.~P.}\ \bibnamefont {Carucci}},
  \bibinfo {author} {\bibfnamefont {S.}~\bibnamefont {Camera}}, \bibinfo
  {author} {\bibfnamefont {M.}~\bibnamefont {Remazeilles}},\ and\ \bibinfo
  {author} {\bibfnamefont {C.}~\bibnamefont {Carbone}},\ }\href
  {https://doi.org/10.48550/arXiv.2509.02644} {\bibinfo {title} {Needlets and
  foreground removal for {{SKAO}} hydrogen intensity maps}} (\bibinfo {year}
  {2025}),\ \Eprint {https://arxiv.org/abs/2509.02644} {arXiv:2509.02644
  [astro-ph]} \BibitemShut {NoStop}%
\bibitem [{\citenamefont {Akrami}\ \emph {et~al.}(2020)\citenamefont {Akrami}
  \emph {et~al.}}]{Planck:2020olo}%
  \BibitemOpen
  \bibfield  {author} {\bibinfo {author} {\bibfnamefont {Y.}~\bibnamefont
  {Akrami}} \emph {et~al.},\ }\bibfield  {title} {\bibinfo {title} {Planck
  intermediate results. {{LVII}}. {{Joint Planck LFI}} and {{HFI}} data
  processing},\ }\href {https://doi.org/10.1051/0004-6361/202038073} {\bibfield
   {journal} {\bibinfo  {journal} {Astron. Astrophys.}\ }\textbf {\bibinfo
  {volume} {643}},\ \bibinfo {pages} {A42} (\bibinfo {year}
  {2020})}\BibitemShut {NoStop}%
\bibitem [{\citenamefont {McCarthy}\ and\ \citenamefont
  {Hill}(2023)}]{McCarthy:2023cwg}%
  \BibitemOpen
  \bibfield  {author} {\bibinfo {author} {\bibfnamefont {F.}~\bibnamefont
  {McCarthy}}\ and\ \bibinfo {author} {\bibfnamefont {J.~C.}\ \bibnamefont
  {Hill}},\ }\href {https://doi.org/10.48550/arXiv.2308.16260} {\bibinfo
  {title} {Cross-correlation of the thermal {{Sunyaev--Zel}}'dovich and {{CMB}}
  lensing signals in {{Planck PR4}} data with robust {{CIB}} decontamination}}
  (\bibinfo {year} {2023}),\ \Eprint {https://arxiv.org/abs/2308.16260}
  {arXiv:2308.16260 [astro-ph]} \BibitemShut {NoStop}%
\bibitem [{\citenamefont {Alonso}\ \emph {et~al.}(2019)\citenamefont {Alonso},
  \citenamefont {Sanchez},\ and\ \citenamefont {Slosar}}]{Alonso:2018jzx}%
  \BibitemOpen
  \bibfield  {author} {\bibinfo {author} {\bibfnamefont {D.}~\bibnamefont
  {Alonso}}, \bibinfo {author} {\bibfnamefont {J.}~\bibnamefont {Sanchez}},\
  and\ \bibinfo {author} {\bibfnamefont {A.}~\bibnamefont {Slosar}},\
  }\bibfield  {title} {\bibinfo {title} {A unified pseudo-$c_\ell$ framework},\
  }\href {https://doi.org/10.1093/mnras/stz093} {\bibfield  {journal} {\bibinfo
   {journal} {Monthly Notices of the Royal Astronomical Society}\ }\textbf
  {\bibinfo {volume} {484}},\ \bibinfo {pages} {4127} (\bibinfo {year}
  {2019})},\ \Eprint {https://arxiv.org/abs/1809.09603} {arXiv:1809.09603
  [astro-ph]} \BibitemShut {NoStop}%
\bibitem [{\citenamefont {{Garc{\'i}a-Garc{\'i}a}}\ \emph
  {et~al.}(2019)\citenamefont {{Garc{\'i}a-Garc{\'i}a}}, \citenamefont
  {Alonso},\ and\ \citenamefont {Bellini}}]{Garcia-Garcia:2019bku}%
  \BibitemOpen
  \bibfield  {author} {\bibinfo {author} {\bibfnamefont {C.}~\bibnamefont
  {{Garc{\'i}a-Garc{\'i}a}}}, \bibinfo {author} {\bibfnamefont
  {D.}~\bibnamefont {Alonso}},\ and\ \bibinfo {author} {\bibfnamefont
  {E.}~\bibnamefont {Bellini}},\ }\bibfield  {title} {\bibinfo {title}
  {Disconnected pseudo-{{$C_\ell$}} covariances for projected large-scale
  structure data},\ }\href {https://doi.org/10.1088/1475-7516/2019/11/043}
  {\bibfield  {journal} {\bibinfo  {journal} {Journal of Cosmology and
  Astroparticle Physics}\ }\textbf {\bibinfo {volume} {2019}}\bibfield
  {number} {\bibinfo  {number} { (11)},\ \bibinfo {pages} {043}},\ }\Eprint
  {https://arxiv.org/abs/1906.11765} {arXiv:1906.11765 [astro-ph]} \BibitemShut
  {NoStop}%
\bibitem [{\citenamefont {Nicola}\ \emph {et~al.}(2021)\citenamefont {Nicola},
  \citenamefont {{Garc{\'i}a-Garc{\'i}a}}, \citenamefont {Alonso},
  \citenamefont {Dunkley}, \citenamefont {Ferreira}, \citenamefont {Slosar},\
  and\ \citenamefont {Spergel}}]{Nicola:2020lhi}%
  \BibitemOpen
  \bibfield  {author} {\bibinfo {author} {\bibfnamefont {A.}~\bibnamefont
  {Nicola}}, \bibinfo {author} {\bibfnamefont {C.}~\bibnamefont
  {{Garc{\'i}a-Garc{\'i}a}}}, \bibinfo {author} {\bibfnamefont
  {D.}~\bibnamefont {Alonso}}, \bibinfo {author} {\bibfnamefont
  {J.}~\bibnamefont {Dunkley}}, \bibinfo {author} {\bibfnamefont {P.~G.}\
  \bibnamefont {Ferreira}}, \bibinfo {author} {\bibfnamefont {A.}~\bibnamefont
  {Slosar}},\ and\ \bibinfo {author} {\bibfnamefont {D.~N.}\ \bibnamefont
  {Spergel}},\ }\bibfield  {title} {\bibinfo {title} {Cosmic shear power
  spectra in practice},\ }\href {https://doi.org/10.1088/1475-7516/2021/03/067}
  {\bibfield  {journal} {\bibinfo  {journal} {Journal of Cosmology and
  Astroparticle Physics}\ }\textbf {\bibinfo {volume} {2021}}\bibfield
  {number} {\bibinfo  {number} { (03)},\ \bibinfo {pages} {067}},\ }\Eprint
  {https://arxiv.org/abs/2010.09717} {arXiv:2010.09717 [astro-ph]} \BibitemShut
  {NoStop}%
\bibitem [{\citenamefont {Mesinger}\ \emph {et~al.}(2011)\citenamefont
  {Mesinger}, \citenamefont {Furlanetto},\ and\ \citenamefont
  {Cen}}]{Mesinger:2010ne}%
  \BibitemOpen
  \bibfield  {author} {\bibinfo {author} {\bibfnamefont {A.}~\bibnamefont
  {Mesinger}}, \bibinfo {author} {\bibfnamefont {S.}~\bibnamefont
  {Furlanetto}},\ and\ \bibinfo {author} {\bibfnamefont {R.}~\bibnamefont
  {Cen}},\ }\bibfield  {title} {\bibinfo {title} {21cmfast: A fast,
  seminumerical simulation of the high-redshift 21-cm signal},\ }\href
  {https://doi.org/10.1111/j.1365-2966.2010.17731.x} {\bibfield  {journal}
  {\bibinfo  {journal} {Monthly Notices of the Royal Astronomical Society}\
  }\textbf {\bibinfo {volume} {411}},\ \bibinfo {pages} {955} (\bibinfo {year}
  {2011})}\BibitemShut {NoStop}%
\bibitem [{\citenamefont {Park}\ \emph {et~al.}(2019)\citenamefont {Park},
  \citenamefont {Mesinger}, \citenamefont {Greig},\ and\ \citenamefont
  {Gillet}}]{Park:2018ljd}%
  \BibitemOpen
  \bibfield  {author} {\bibinfo {author} {\bibfnamefont {J.}~\bibnamefont
  {Park}}, \bibinfo {author} {\bibfnamefont {A.}~\bibnamefont {Mesinger}},
  \bibinfo {author} {\bibfnamefont {B.}~\bibnamefont {Greig}},\ and\ \bibinfo
  {author} {\bibfnamefont {N.}~\bibnamefont {Gillet}},\ }\bibfield  {title}
  {\bibinfo {title} {Inferring the astrophysics of reionization and cosmic dawn
  from galaxy luminosity functions and the 21-cm signal},\ }\href
  {https://doi.org/10.1093/mnras/stz032} {\bibfield  {journal} {\bibinfo
  {journal} {Monthly Notices of the Royal Astronomical Society}\ }\textbf
  {\bibinfo {volume} {484}},\ \bibinfo {pages} {933} (\bibinfo {year}
  {2019})}\BibitemShut {NoStop}%
\bibitem [{\citenamefont {Murray}\ \emph {et~al.}(2020)\citenamefont {Murray},
  \citenamefont {Greig}, \citenamefont {Mesinger}, \citenamefont {Mu{\~n}oz},
  \citenamefont {Qin}, \citenamefont {Park},\ and\ \citenamefont
  {Watkinson}}]{Murray:2020trn}%
  \BibitemOpen
  \bibfield  {author} {\bibinfo {author} {\bibfnamefont {S.~G.}\ \bibnamefont
  {Murray}}, \bibinfo {author} {\bibfnamefont {B.}~\bibnamefont {Greig}},
  \bibinfo {author} {\bibfnamefont {A.}~\bibnamefont {Mesinger}}, \bibinfo
  {author} {\bibfnamefont {J.~B.}\ \bibnamefont {Mu{\~n}oz}}, \bibinfo {author}
  {\bibfnamefont {Y.}~\bibnamefont {Qin}}, \bibinfo {author} {\bibfnamefont
  {J.}~\bibnamefont {Park}},\ and\ \bibinfo {author} {\bibfnamefont {C.~A.}\
  \bibnamefont {Watkinson}},\ }\bibfield  {title} {\bibinfo {title}
  {{{21cmFAST}} v3: {{A Python-integrated C}} code for generating {{3D}}
  realizations of the cosmic 21cm signal.},\ }\href
  {https://doi.org/10.21105/joss.02582} {\bibfield  {journal} {\bibinfo
  {journal} {Journal of Open Source Software}\ }\textbf {\bibinfo {volume}
  {5}},\ \bibinfo {pages} {2582} (\bibinfo {year} {2020})}\BibitemShut
  {NoStop}%
\bibitem [{\citenamefont {Harris}\ \emph {et~al.}(2020)\citenamefont {Harris}
  \emph {et~al.}}]{Harris:2020xlr}%
  \BibitemOpen
  \bibfield  {author} {\bibinfo {author} {\bibfnamefont {C.~R.}\ \bibnamefont
  {Harris}} \emph {et~al.},\ }\bibfield  {title} {\bibinfo {title} {Array
  programming with {{NumPy}}},\ }\href
  {https://doi.org/10.1038/s41586-020-2649-2} {\bibfield  {journal} {\bibinfo
  {journal} {Nature}\ }\textbf {\bibinfo {volume} {585}},\ \bibinfo {pages}
  {357} (\bibinfo {year} {2020})}\BibitemShut {NoStop}%
\bibitem [{\citenamefont {Virtanen}\ \emph {et~al.}(2020)\citenamefont
  {Virtanen} \emph {et~al.}}]{Virtanen:2019joe}%
  \BibitemOpen
  \bibfield  {author} {\bibinfo {author} {\bibfnamefont {P.}~\bibnamefont
  {Virtanen}} \emph {et~al.},\ }\bibfield  {title} {\bibinfo {title} {{{SciPy}}
  1.0: Fundamental algorithms for scientific computing in {{Python}}},\ }\href
  {https://doi.org/10.1038/s41592-019-0686-2} {\bibfield  {journal} {\bibinfo
  {journal} {Nature Methods}\ }\textbf {\bibinfo {volume} {17}},\ \bibinfo
  {pages} {261} (\bibinfo {year} {2020})}\BibitemShut {NoStop}%
\bibitem [{\citenamefont {Hunter}(2007)}]{Hunter:2007ouj}%
  \BibitemOpen
  \bibfield  {author} {\bibinfo {author} {\bibfnamefont {J.~D.}\ \bibnamefont
  {Hunter}},\ }\bibfield  {title} {\bibinfo {title} {Matplotlib: {{A 2D
  Graphics Environment}}},\ }\href {https://doi.org/10.1109/MCSE.2007.55}
  {\bibfield  {journal} {\bibinfo  {journal} {Computing in Science \&
  Engineering}\ }\textbf {\bibinfo {volume} {9}},\ \bibinfo {pages} {90}
  (\bibinfo {year} {2007})}\BibitemShut {NoStop}%
\bibitem [{\citenamefont {Robitaille}\ \emph {et~al.}(2013)\citenamefont
  {Robitaille} \emph {et~al.}}]{Astropy:2013muo}%
  \BibitemOpen
  \bibfield  {author} {\bibinfo {author} {\bibfnamefont {T.~P.}\ \bibnamefont
  {Robitaille}} \emph {et~al.} (\bibinfo {collaboration} {The Astropy
  Collaboration}),\ }\bibfield  {title} {\bibinfo {title} {Astropy: {{A}}
  community {{Python}} package for astronomy},\ }\href
  {https://doi.org/10.1051/0004-6361/201322068} {\bibfield  {journal} {\bibinfo
   {journal} {Astronomy \& Astrophysics}\ }\textbf {\bibinfo {volume} {558}},\
  \bibinfo {pages} {A33} (\bibinfo {year} {2013})}\BibitemShut {NoStop}%
\bibitem [{\citenamefont {{Price-Whelan}}\ \emph {et~al.}(2018)\citenamefont
  {{Price-Whelan}} \emph {et~al.}}]{Astropy:2018wqo}%
  \BibitemOpen
  \bibfield  {author} {\bibinfo {author} {\bibfnamefont {A.~M.}\ \bibnamefont
  {{Price-Whelan}}} \emph {et~al.} (\bibinfo {collaboration} {The Astropy
  Collaboration}),\ }\bibfield  {title} {\bibinfo {title} {The {{Astropy
  Project}}: {{Building}} an inclusive, open-science project and status of the
  v2.0 core package},\ }\href {https://doi.org/10.3847/1538-3881/aabc4f}
  {\bibfield  {journal} {\bibinfo  {journal} {The Astronomical Journal}\
  }\textbf {\bibinfo {volume} {156}},\ \bibinfo {pages} {123} (\bibinfo {year}
  {2018})},\ \Eprint {https://arxiv.org/abs/1801.02634} {arXiv:1801.02634
  [astro-ph]} \BibitemShut {NoStop}%
\bibitem [{\citenamefont {{Price-Whelan}}\ \emph {et~al.}(2022)\citenamefont
  {{Price-Whelan}} \emph {et~al.}}]{Astropy:2022ucr}%
  \BibitemOpen
  \bibfield  {author} {\bibinfo {author} {\bibfnamefont {A.~M.}\ \bibnamefont
  {{Price-Whelan}}} \emph {et~al.} (\bibinfo {collaboration} {The Astropy
  Collaboration}),\ }\bibfield  {title} {\bibinfo {title} {The {{Astropy
  Project}}: {{Sustaining}} and {{Growing}} a {{Community-oriented Open-source
  Project}} and the {{Latest Major Release}} (v5.0) of the {{Core Package}}},\
  }\href {https://doi.org/10.3847/1538-4357/ac7c74} {\bibfield  {journal}
  {\bibinfo  {journal} {The Astrophysical Journal}\ }\textbf {\bibinfo {volume}
  {935}},\ \bibinfo {pages} {167} (\bibinfo {year} {2022})},\ \Eprint
  {https://arxiv.org/abs/2206.14220} {arXiv:2206.14220 [astro-ph]} \BibitemShut
  {NoStop}%
\bibitem [{\citenamefont {Murray}\ \emph {et~al.}(2013)\citenamefont {Murray},
  \citenamefont {Power},\ and\ \citenamefont {Robotham}}]{Murray:2013qza}%
  \BibitemOpen
  \bibfield  {author} {\bibinfo {author} {\bibfnamefont {S.}~\bibnamefont
  {Murray}}, \bibinfo {author} {\bibfnamefont {C.}~\bibnamefont {Power}},\ and\
  \bibinfo {author} {\bibfnamefont {A.}~\bibnamefont {Robotham}},\ }\href
  {https://doi.org/10.48550/arXiv.1306.6721} {\bibinfo {title} {{{HMFcalc}}:
  {{An Online Tool}} for {{Calculating Dark Matter Halo Mass Functions}}}}
  (\bibinfo {year} {2013}),\ \Eprint {https://arxiv.org/abs/1306.6721}
  {arXiv:1306.6721 [astro-ph]} \BibitemShut {NoStop}%
\bibitem [{\citenamefont {Murray}\ \emph {et~al.}(2021)\citenamefont {Murray},
  \citenamefont {Diemer}, \citenamefont {Chen}, \citenamefont {Neuhold},
  \citenamefont {Schnapp}, \citenamefont {Peruzzi}, \citenamefont {Blevins},\
  and\ \citenamefont {Engelman}}]{Murray:2020dcd}%
  \BibitemOpen
  \bibfield  {author} {\bibinfo {author} {\bibfnamefont {S.~G.}\ \bibnamefont
  {Murray}}, \bibinfo {author} {\bibfnamefont {B.}~\bibnamefont {Diemer}},
  \bibinfo {author} {\bibfnamefont {Z.}~\bibnamefont {Chen}}, \bibinfo {author}
  {\bibfnamefont {A.~G.}\ \bibnamefont {Neuhold}}, \bibinfo {author}
  {\bibfnamefont {M.~A.}\ \bibnamefont {Schnapp}}, \bibinfo {author}
  {\bibfnamefont {T.}~\bibnamefont {Peruzzi}}, \bibinfo {author} {\bibfnamefont
  {D.}~\bibnamefont {Blevins}},\ and\ \bibinfo {author} {\bibfnamefont
  {T.}~\bibnamefont {Engelman}},\ }\bibfield  {title} {\bibinfo {title}
  {{{TheHaloMod}}: {{An}} online calculator for the halo model},\ }\href
  {https://doi.org/10.1016/j.ascom.2021.100487} {\bibfield  {journal} {\bibinfo
   {journal} {Astron. Comput.}\ }\textbf {\bibinfo {volume} {36}},\ \bibinfo
  {pages} {100487} (\bibinfo {year} {2021})}\BibitemShut {NoStop}%
\bibitem [{\citenamefont {G{\'o}rski}\ \emph {et~al.}(2005)\citenamefont
  {G{\'o}rski}, \citenamefont {Hivon}, \citenamefont {Banday}, \citenamefont
  {Wandelt}, \citenamefont {Hansen}, \citenamefont {Reinecke},\ and\
  \citenamefont {Bartelman}}]{Gorski:2004by}%
  \BibitemOpen
  \bibfield  {author} {\bibinfo {author} {\bibfnamefont {K.~M.}\ \bibnamefont
  {G{\'o}rski}}, \bibinfo {author} {\bibfnamefont {E.}~\bibnamefont {Hivon}},
  \bibinfo {author} {\bibfnamefont {A.~J.}\ \bibnamefont {Banday}}, \bibinfo
  {author} {\bibfnamefont {B.~D.}\ \bibnamefont {Wandelt}}, \bibinfo {author}
  {\bibfnamefont {F.~K.}\ \bibnamefont {Hansen}}, \bibinfo {author}
  {\bibfnamefont {M.}~\bibnamefont {Reinecke}},\ and\ \bibinfo {author}
  {\bibfnamefont {M.}~\bibnamefont {Bartelman}},\ }\bibfield  {title} {\bibinfo
  {title} {{{HEALPix}} - {{A Framework}} for high resolution discretization,
  and fast analysis of data distributed on the sphere},\ }\href
  {https://doi.org/10.1086/427976} {\bibfield  {journal} {\bibinfo  {journal}
  {Astrophys. J.}\ }\textbf {\bibinfo {volume} {622}},\ \bibinfo {pages} {759}
  (\bibinfo {year} {2005})}\BibitemShut {NoStop}%
\bibitem [{\citenamefont {Zonca}\ \emph {et~al.}(2019)\citenamefont {Zonca},
  \citenamefont {Singer}, \citenamefont {Lenz}, \citenamefont {Reinecke},
  \citenamefont {Rosset}, \citenamefont {Hivon},\ and\ \citenamefont
  {Gorski}}]{Zonca:2019vzt}%
  \BibitemOpen
  \bibfield  {author} {\bibinfo {author} {\bibfnamefont {A.}~\bibnamefont
  {Zonca}}, \bibinfo {author} {\bibfnamefont {L.}~\bibnamefont {Singer}},
  \bibinfo {author} {\bibfnamefont {D.}~\bibnamefont {Lenz}}, \bibinfo {author}
  {\bibfnamefont {M.}~\bibnamefont {Reinecke}}, \bibinfo {author}
  {\bibfnamefont {C.}~\bibnamefont {Rosset}}, \bibinfo {author} {\bibfnamefont
  {E.}~\bibnamefont {Hivon}},\ and\ \bibinfo {author} {\bibfnamefont
  {K.}~\bibnamefont {Gorski}},\ }\bibfield  {title} {\bibinfo {title} {Healpy:
  Equal area pixelization and spherical harmonics transforms for data on the
  sphere in {{Python}}},\ }\href {https://doi.org/10.21105/joss.01298}
  {\bibfield  {journal} {\bibinfo  {journal} {Journal of Open Source Software}\
  }\textbf {\bibinfo {volume} {4}},\ \bibinfo {pages} {1298} (\bibinfo {year}
  {2019})}\BibitemShut {NoStop}%
\bibitem [{\citenamefont {Baker}\ and\ \citenamefont
  {Liu}(2026{\natexlab{b}})}]{Baker_code_repo}%
  \BibitemOpen
  \bibfield  {author} {\bibinfo {author} {\bibfnamefont {E.}~\bibnamefont
  {Baker}}\ and\ \bibinfo {author} {\bibfnamefont {H.}~\bibnamefont {Liu}},\
  }\href {https://doi.org/10.5281/zenodo.21222684} {\bibinfo {title} {Dark
  {{Photons}} in the {{Radio Sky}} [{{Software}}]}},\ \bibinfo {howpublished}
  {Zenodo} (\bibinfo {year} {2026}{\natexlab{b}})\BibitemShut {NoStop}%
\end{thebibliography}%

\end{document}